\documentclass[11pt]{article}
\usepackage{jheppub}
\usepackage{amsmath,amssymb,amsfonts,graphicx,slashed,color, mathtools, amsthm}
\usepackage{comment}
\usepackage{enumerate}
\usepackage{float}
\usepackage{caption}
\usepackage{subcaption}
\usepackage{pdflscape}
\usepackage{physics}
\usepackage{tikz}
\usepackage{mathrsfs}
\usepackage{amsthm}
\usetikzlibrary{arrows.meta, decorations.markings}
\tikzset{
    pil/.style={->, thick, shorten <=2pt, shorten >=2pt,},
    mid arrow/.style={postaction={decorate,decoration={
        markings,
        mark=at position .5 with {\arrow[#1]{stealth'}}
    }}},
}

\newtheorem{theorem}{Theorem}
\newtheorem{proposal}[theorem]{Proposal}
\newtheorem{proposition}[theorem]{Proposition}

\newcommand{\be}{\begin{equation}}
\newcommand{\ee}{\end{equation}}
\newcommand{\bea}{\begin{eqnarray}}
\newcommand{\eea}{\end{eqnarray}}
\newcommand{\beas}{\begin{eqnarray*}}
\newcommand{\eeas}{\end{eqnarray*}}
\newcommand{\ba}{\begin{array}}
\newcommand{\ea}{\end{array}}

\newcommand{\lads}{L_{\textnormal{AdS}}}

\newcommand{\V}{\hat{\mathcal{V}}}
\newcommand{\W}{\hat{\mathcal{W}}}
\newcommand{\X}{\hat{\mathcal{X}}}
\newcommand{\Y}{\hat{\mathcal{Y}}}

\title{On sufficient conditions for holographic scattering}

\author[a, b, c]{Caroline Lima,}
\author[a]{Sabrina Pasterski,}
\author[a]{and Chris Waddell}

\affiliation[a]{Perimeter Institute for Theoretical Physics, 31 Caroline St N, Waterloo, ON\ N2L 2Y5, Canada}
\affiliation[b]{Department of Physics and Astronomy, University of Waterloo, Waterloo, Ontario N2L 3G1, Canada}
\affiliation[c]{Institute for Quantum Computing, University of Waterloo, Waterloo, Ontario, N2L 3G1, Canada}

\emailAdd{clima@perimeterinstitute.ca, spasterski@perimeterinstitute.ca, cwaddell@perimeterinstitute.ca}

\abstract{Holography implies scattering in the bulk can be mediated by entanglement on the boundary. The connected wedge theorem (CWT) of May, Penington, and Sorce is a concrete example where bulk scattering implies correlation between certain boundary regions. However the converse does not hold. We investigate a recent proposal of Leutheusser and Liu for a generalization of the CWT with converse. We prove the forward direction: having pairs of CFT ``input'' (and likewise ``output'') regions in a phase with connected entanglement wedge implies that a particular bulk subregion (the intersection of ``input'' and ``output'' entanglement wedges) is non-empty. We then establish a modified version of the proposal which has a converse, and identify counter-examples to the stronger conjecture. }
\keywords{}

\begin{document}

\maketitle

\section{Introduction} \label{sec:intro}

The connected wedge theorem (CWT) of May, Penington, and Sorce \cite{May:2019yxi, May:2019odp, May:2021nrl} suggests that extending quantum information theoretic considerations into the relativistic regime can put novel constraints on gravity. 
The nature of these constraints is interesting: they are contingent on the existence of certain ``scattering regions'' in the bulk spacetime, which can be given an explicit operational definition from the CFT perspective. The question of what conditions on ``holographic'' quantum states can ensure the emergence of a local bulk is therefore naturally translated to a question about which quantum states offer the appropriate resources to perform certain distributed computing tasks.

In the context of the CWT, one obtains a condition on the state of the CFT which is necessary (though not sufficient) for the emergence of a bulk scattering region. The basic idea is illustrated in figure \ref{fig:CWT}. We imagine that Alice and Bob have access to CFT regions $\hat{\mathcal{V}}_{1}$ and $\hat{\mathcal{V}}_{2}$ respectively. At $c_{1}$, Alice receives some unknown state $| \psi \rangle$ of a qubit, while at $c_{2}$, Bob receives some bit $i$.\footnote{The precise information processing task we are describing here is not exactly the same as the $\textbf{B}_{84}$ task considered in \cite{May:2019yxi, May:2019odp, May:2021nrl}, though it is similar in spirit.} Though they are restricted to act within mutually spacelike spacetime regions, their goal is to send the state $| \psi \rangle$ to the output location $r_{i}$. The causal structure in the CFT seems to prevent Alice from ``deciding'' which output point to send her state to before receiving information about the desired destination from Bob, and the no-cloning requirement prevents her from sending the state to both output points. Nevertheless, if it is possible to transmit information from $\hat{\mathcal{V}}_{1}$ and $\hat{\mathcal{V}}_{2}$ through the bulk to some bulk ``scattering point'', and from this scattering point to the regions $\hat{\mathcal{W}}_{1}$ and $\hat{\mathcal{W}}_{2}$, then it is clear that Alice and Bob must be able to complete their task through some judicious choice of operations. The resolution of this apparent tension is that one can also construct a non-local boundary protocol to complete the task, but it requires large mutual information $I(\hat{\mathcal{V}}_{1}:\hat{\mathcal{V}}_{2}) = O(1/G)$, i.e. a connected entanglement wedge for $\hat{\mathcal{V}}_{1} \cup \hat{\mathcal{V}}_{2}$.

\begin{figure}
    \centering   
    \includegraphics[height=6cm]{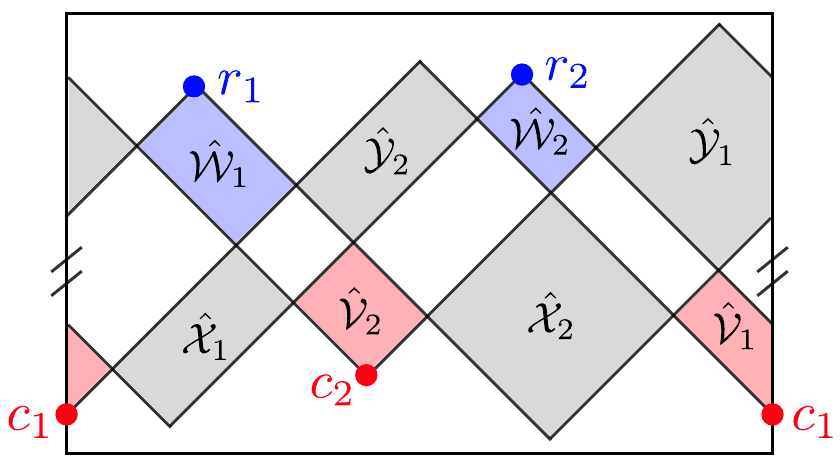}
    \caption{Boundary geometrical set-up for the connected wedge theorem. The rectangle represents a fixed time interval of the CFT; since this is on a cylinder, the left and right edges are identified. Input points $c_{1}, c_{2}$ are red, while output points $r_{1}, r_{2}$ are blue. The decision regions $\hat{\mathcal{V}}_{1}, \hat{\mathcal{V}}_{2}$ are shaded light red, while the decision regions $\hat{\mathcal{W}}_{1}, \hat{\mathcal{W}}_{2}$ are shaded light blue. The boundary scattering region $\hat{\mathcal{S}} \equiv \hat{\mathcal{J}}^{+}[c_{1}] \cap \hat{\mathcal{J}}^{+}[c_{2}] \cap \hat{\mathcal{J}}^{-}[r_{1}] \cap \hat{\mathcal{J}}^{-}[r_{2}]$ is empty. }
\end{figure}\label{fig:CWT}

As we have emphasized, the statement of the CWT is an implication, rather than an equivalence, relating boundary correlation to bulk scattering. On the other hand, formulating an ``if and only if'' statement relating these notions would be useful for understanding how bulk locality is related to the entanglement structure of quantum states. There have recently been two different approaches to the question of a CWT-with-converse in the literature. One of these, appearing in \cite{Caminiti:2024ctd, Caminiti:2025gyv}, 
has investigated whether there are extra conditions, in addition to connectedness of the entanglement wedge, which can guarantee scattering in the bulk. 
In \cite{Caminiti:2024ctd}, it was found within certain AdS$_{3}$ quotient spacetimes corresponding to pure CFT states that a necessary and sufficient criterion for scattering is that, in addition to a connected entanglement wedge, one should have a particular relationship between the areas of extremal but possibly non-minimal surfaces.\footnote{This condition was still not sufficient for scattering in the case of mixed states, as observed by \cite{Caminiti:2024ctd} in the BTZ black hole spacetime.} This was interpreted (invoking \cite{Balasubramanian:2014sra}) as requiring a particular entanglement pattern between the internal degrees of freedom in the subregions with large mutual information, connecting to earlier work relating entanglement structure of internal degrees of freedom to sub-AdS locality and bulk scattering (e.g. \cite{Susskind:1998vk, Heemskerk:2009pn}).

Alternatively, a generalized connected wedge (GCW) proposal has been conjectured by Leutheusser and Liu in \cite{Leutheusser:2024yvf}. They defined a very different bulk subregion
\begin{equation} \label{eq:def_SE}
    \mathcal{S}_{E} \equiv \mathcal{E}[\V_{1} \cup \V_{2}] \cap \mathcal{E}[\W_{1} \cup \W_{2}] \: ,
\end{equation}
where $\mathcal{E}[\cdot]$ denotes the entanglement wedge, which no longer admits a simple interpretation as a scattering region, but which can nevertheless be shown to contain the scattering region $\mathcal{S}$ appearing in the standard CWT.\footnote{This is an immediate consequence of the fact that $\mathcal{S} \subset \mathcal{E}[\V_{1} \cup \V_{2}]$, as shown in \cite{May:2019odp}, and the result $\mathcal{S} \subset \mathcal{E}[\W_{1} \cup \W_{2}]$ which follows from time-reversal.} They argued that the non-emptiness of $\mathcal{S}_{E}$ should be equivalent to the requirement $I(\V_{1} : \V_{2}) = O(1/G)$. In addition to being a version of the CWT with a converse, the conjecture also has the benefit that it can be interpreted as an explicit statement about properties of von Neumann algebras in large $N$ theories: it says that the super-additivity property\footnote{Recall that the algebraic union $\mathcal{A}_{1} \vee \mathcal{A}_{2} \equiv (\mathcal{A}_{1} \cup \mathcal{A}_{2})''$, with $'$ denoting the commutant, is the smallest von Neumann algebra containing both $\mathcal{A}_{1}$ and $\mathcal{A}_{2}$. }
\begin{equation}
    \mathcal{A}_{\V_{1} \cup \V_{2}} \neq \mathcal{A}_{\V_{1}} \vee \mathcal{A}_{\V_{2}}
\end{equation}
of the algebras associated with the entanglement wedges of $\V_{1}$, $\V_{2}$, and $\V_{1} \cup \V_{2}$ is equivalent to the requirement that one has a non-empty intersection of algebras associated to the entanglement wedges of $\V_{1} \cup \V_{2}$ and $\W_{1} \cup \W_{2}$, 
\begin{equation}
    \mathcal{A}_{\hat{\mathcal{V}}_{1} \cup \hat{\mathcal{V}}_{2}} \cap \mathcal{A}_{\hat{\mathcal{W}}_{1} \cup \hat{\mathcal{W}}_{2}} \neq \emptyset \: .
\end{equation}
This alternative approach to a theorem-with-converse is not mutually exclusive with the approach in \cite{Caminiti:2024ctd}, though it has a very different flavour; for example, the GCW proposal is apparently insensitive to the specific pattern of entanglement between ``internal space'' degrees of freedom 
emphasized in \cite{Caminiti:2024ctd}. 
Moreover, the information theoretic interpretation of the GCW proposal is necessarily quite different from that of the ordinary CWT, since $\mathcal{S}_{E}$ is not truly a scattering region, but is rather a bulk subregion accessible to an agent with access to \textit{both} $\V_{1}$ and $\V_{2}$ (and to an agent with access to \textit{both} $\W_{1}$ and $\W_{2}$). 

In this note, we will show that the GCW proposal described above does not hold. Thinking about this proposal as a statement relating (1) the connectedness of entanglement wedges for unions of ``input regions'' $\hat{\mathcal{V}}_{1}$, $\hat{\mathcal{V}}_{2}$ and/or ``output regions'' $\hat{\mathcal{W}}_{1}$, $\hat{\mathcal{W}}_{2}$ and (2) emptiness of $\mathcal{S}_{E}$, we show that the only possible logical relationship between these conditions is that having both $\mathcal{E}[ \V_{1} \cup \V_{2}]$ and $\mathcal{E}[\W_{1} \cup \W_{2}]$ connected implies $\mathcal{S}_{E} \neq \emptyset$, and we prove this implication in generality 
using the methods of \cite{May:2019odp}, subject to analogous assumptions. 
A slightly stronger version of this implication is that the non-emptiness of the bulk region
\begin{equation}\label{se2}
    \tilde{\mathcal{S}}_{E} \equiv \left( \mathcal{E}[\V_{1} \cup \V_{2}] \setminus (\mathcal{E}[\V_{1}] \cup \mathcal{E}[\V_{2}]) \right) \cap \left( \mathcal{E}[\W_{1} \cup \W_{2}] \setminus (\mathcal{E}[\W_{1}] \cup \mathcal{E}[\W_{2}]) \right)
\end{equation}
is equivalent to having large mutual informations $I(\V_{1} : \V_{2}) = O(1/G)$ \textit{and} $I(\W_{1} : \W_{2}) = O(1/G)$.\footnote{We note that, despite the fact that specifying $\V_{1}, \V_{2}$ determines $\W_{1}, \W_{2}$, $I(\V_{1} : \V_{2}) = O(1/G)$ does not imply $I(\W_{1} : \W_{2}) = O(1/G)$, and vice versa. See section \ref{sec:counter-ex} for examples.} This can be viewed as a kind of connected wedge theorem-with-converse, equating the existence of a certain bulk region to large boundary correlation, albeit with one direction enforced by fiat in the definition of $\tilde{\mathcal{S}}_{E}$. 

While our results show that the proposed theorem-with-converse furnished by the GCW is not true, the conclusion that we derive is nevertheless in the opposite direction from the standard CWT: whereas the CWT states that large boundary correlation is necessary for the existence of a certain bulk region, our finding is that large boundary correlation is sufficient to ensure the existence of a (different) bulk spacetime region. This is notable, given that we are not specifying any more detailed features of the entanglement structure, and it is interesting to consider whether this spacetime region has an operational interpretation from the boundary point of view; we comment on the information theoretic interpretation in the discussion. 

As a byproduct of our discussion, we note that, in addition to the lower bound on the mutual information in terms of the ``ridge'' of the usual scattering region $\mathcal{S}$ established in \cite{May:2019odp}, one also obtains an \textit{upper bound} on the mutual information from an analogous ``ridge'' of the region $\mathcal{S}_{E}$.\footnote{In particular, we describe in section \ref{sec:mi_bound} an upper bound and a definition of the ridge applicable when the intersection of backward null sheets from the boundary regions $\Y_{i}$ does not dip below the maximin slice associated with regions $\V_{i}$. A more general analysis will be provided in \cite{ap}. }
This provides an alternative perspective on the implication that $\mathcal{E}[\V_{1} \cup \V_{2}]$, $\mathcal{E}[\W_{1} \cup \W_{2}]$ 
connected ensure that $\mathcal{S}_{E}$ is non-empty, by showing that a notion of size for this region is lower bounded by the mutual information.

This paper is organized as follows. In section~\ref{sec:review} we review the connected wedge theorem of~\cite{May:2019yxi, May:2019odp, May:2021nrl} as well as the proposal from~\cite{Leutheusser:2024yvf} of a generalized connected wedge theorem with converse. We then establish the forward direction for (a small modification of) the latter proposal in section~\ref{sec:proof}, and show that the entanglement wedge intersections give an upper bound on the mutual information in section~\ref{sec:mi_bound} before establishing a weaker GCW theorem-with-converse in section~\ref{sec:converse}. Finally, we demonstrate counter-examples to the converse direction of the original GCW proposal in section~\ref{sec:counter-ex}. 
We close with a discussion of the boundary interpretations of our results in section \ref{sec:discussion}. Some additional proof details and conical defect computations are included in the appendix.
Before proceeding let us establish some notation and conventions.

\paragraph{Note added:} The day before this work  appeared on the arXiv, the paper \cite{Zhao:2025sso} was posted, reaching a similar conclusion to our proposition \ref{GCWT2}. The original version of \cite{Zhao:2025sso} also asserted a converse statement, which is ruled out by the counter-examples presented in our work. This assertion was revised in v2 of \cite{Zhao:2025sso}, whose conclusions are now consistent with ours.

\subsection{Notation and conventions}

In the remainder, we will make use of the following notation, conventions, and definitions for spacetime regions:
\begin{itemize}
    \item Bulk spacetime regions are denoted with capital script Latin letters, $\mathcal{U}, \: \mathcal{V}, \: \mathcal{W} , \: \ldots$ We will add a hat to take the restriction to the conformal boundary, $\hat{\mathcal{U}} , \: \V, \: \W, \: \ldots$
    \item The causal future of a spacetime region $\mathcal{A}$ is denoted by $\mathcal{J}^+[\mathcal{A}]$; the causal past of a spacetime region $\mathcal{A}$ is denoted by $\mathcal{J}^-[\mathcal{A}]$.
    \item The domain of dependence of a region $\mathcal{A}$ is defined as the set of points $p$ such that every inextendible causal curve through $p$ intersects $\mathcal{A}$. We denote the domain of dependence by $\mathcal{D}[\mathcal{A}]$. The future/past domains of dependence, i.e. the portions of $\mathcal{D}[\mathcal{A}]$ lying to the future/past of $\mathcal{A}$, will be denoted by $\mathcal{D}^{+}[\mathcal{A}]$ and $\mathcal{D}^{-}[\mathcal{A}]$ respectively.
    \item The spacelike complement of a region $\mathcal{A}$ is denoted by $\mathcal{A}'$; the set theoretic complement will instead be denoted by $\mathcal{A}^{c}$. 
    The 
    boundary of a region $\mathcal{A}$ is denoted by $\partial \mathcal{A}$. The interior of $\mathcal{A}$ will be denoted by $\text{int}[\mathcal{A}]$. 
    \item The causal wedge \cite{Hubeny:2012wa, Bousso:2012mh} associated to a boundary region $\hat{\mathcal{A}}$ will be denoted by $\mathcal{C}[\hat{\mathcal{A}}]$, and the entanglement wedge \cite{Czech:2012bh, Headrick:2014cta, Wall:2012uf} associated to $\hat{\mathcal{A}}$ will be denoted by $\mathcal{E}[\hat{\mathcal{A}}]$. Note that these are codimension-0 spacetime regions. 
\end{itemize}

\section{Review: The CWT and the GCW proposal} \label{sec:review}

The CWT hinges on assumptions about bulk and boundary causal structure, so we will begin by defining the spacetime subregions in the bulk and boundary that we will use to characterize this structure, following \cite{May:2019yxi, May:2019odp}. 

\subsubsection*{Boundary and bulk scattering regions}

Let $\mathcal{M}$ be an asymptotically globally AdS spacetime, with a conformal boundary, denoted by $\partial \mathcal{M}$, conformally equivalent to the Lorentzian cylinder; we will use $\overline{\mathcal{M}}$ to denote the union $\mathcal{M} \cup \partial \mathcal{M}$. 
Let $c_{1}, c_{2}$ be two spacelike separated points on the $\partial \mathcal{M}$ which we will refer to as \textit{input points}, and $r_{1}, r_{2}$ be two spacelike separated points on $\partial \mathcal{M}$ which we will refer to as \textit{output points}; see figure \ref{fig:CWT}. The \textit{decision regions} $\V_{i}$ and $\W_{i}$ are then defined by
\begin{equation}
    \V_{i} \equiv \hat{\mathcal{J}}^{+}[c_{i}] \cap \hat{\mathcal{J}}^{-}[r_{1}] \cap \hat{\mathcal{J}}^{-}[r_{2}] \: , \quad \W_{i} \equiv \hat{\mathcal{J}}^{+}[c_{1}] \cap \hat{\mathcal{J}}^{+}[c_{2}] \cap \hat{\mathcal{J}}^{-}[r_{i}] \: . 
\end{equation}
We will be interested in configurations of $c_{1}, c_{2}, r_{1}, r_{2}$ for which all of these regions are non-empty, which in particular requires $r_{i}$ to be in the causal future of $c_{j}$ for all $i, j \in \{1, 2\}$, but for which the \textit{boundary scattering region}
\begin{equation}
    \hat{\mathcal{S}} \equiv \hat{\mathcal{J}}^{+}[c_{1}] \cap \hat{\mathcal{J}}^{+}[c_{2}] \cap \hat{\mathcal{J}}^{-}[r_{1}] \cap \hat{\mathcal{J}}^{-}[r_{2}]
\end{equation}
is empty. This means that it is not possible for information to propagate causally from both $c_{1}$ and $c_{2}$ to some scattering location and then to both $r_{1}$ and $r_{2}$ if this process is constrained to lie entirely within $\partial \mathcal{M}$. 

On the other hand, we will be interested in situations where such causal information processing from $c_{1}, c_{2}$ to $r_{1}, r_{2}$ is possible in the bulk. 
In particular, defining the \textit{(bulk) scattering region}
\begin{equation}
    \mathcal{S} \equiv \mathcal{J}^{+}[c_{1}] \cap \mathcal{J}^{+}[c_{2}] \cap \mathcal{J}^{-}[r_{1}] \cap \mathcal{J}^{-}[r_{2}] \: ,
\end{equation}
we will be interested in situations where $\mathcal{S}$ may be non-empty despite the emptiness of $\hat{\mathcal{S}}$. This would mean that it is possible for information to propagate causally from the input points to some bulk scattering location and then to the output points.

\subsection{The connected wedge theorem}

Having established the relevant notation, we can somewhat informally state the CWT.
\begin{theorem} \label{thm:CWT}
(May-Penington-Sorce  \cite{May:2019odp}) Let $c_{1}, c_{2}, r_{1}, r_{2}$ be a ``bulk-only'' scattering configuration, meaning that $\hat{\mathcal{S}} = \emptyset$ while $\mathcal{S} \neq \emptyset$. Then the entanglement wedge $\mathcal{E}(\V_{1} \cup \V_{2})$ is connected. 
\end{theorem}

\noindent
The implicit assumptions here, required in the proof of the theorem, are:\footnote{We believe there is an additional independent assumption, which is that AdS-Cauchy slices are simply connected. This ensures that e.g. for $r \in \partial \mathcal{M}$, the intersection of $\partial \mathcal{J}^{-}[r]$ with such a slice is connected, which is assumed in the proof of the theorem. We will make a similar assumption in the proofs of section \ref{sec:proof}.}
\begin{itemize}
    \item The null energy condition holds in the bulk spacetime.\footnote{Recall that the null energy condition states that $T_{\mu \nu} k^{\mu} k^{\nu} \geq 0$ for every future-pointing null vector field $k^{\mu}$ in $\mathcal{M}$. This is a requirement for passing from the Raychaudhuri equation to the focusing theorem.} 
    \item The ``maximin'' surface \cite{Wall:2012uf} for $\V_{1} \cup \V_{2}$ exists (and is therefore equal to the HRT surface). 
    \item The spacetime is AdS-hyperbolic (such that $\overline{\mathcal{M}}$ admits a Cauchy slice). 
\end{itemize}
Let us make a few comments about this result.

\subsubsection*{Quantum information theoretic interpretation}

While theorem \ref{thm:CWT} can be proven using geometrical techniques (i.e. the focusing theorem) in general relativity, the statement was motivated by applying quantum information theory considerations to the dual CFT. Specifically, one imagines embedding a cryptographic task known as the $\textbf{B}_{84}$ task in the CFT. In this task, Alice receives some random classical bit $q \in \{0, 1\}$ at input point $c_{1}$, Bob receives a quantum state $H^{q} | b \rangle$ (with $H$ the Hadamard gate) involving some independent random bit $b \in \{0, 1\}$ at input point $c_{2}$, and they must coordinate to make the value of $b$ available at both $r_{1}$ and $r_{2}$. In the obvious local strategy to complete this task, $q$ and $H^{q}|b \rangle$ can be brought to the same location to decode the state, at which point $|b \rangle$ can be measured in the computational basis so that $b$ can be sent to the output points. Because the task can apparently be completed (exploiting the emergent bulk) in the absence of a local strategy in the CFT for bulk-only scattering configurations, Alice and Bob must employ a non-local strategy from the boundary perspective, which turns out to require shared correlation \cite{Tomamichel_2013}. Indeed, considering the $\textbf{B}_{84}^{\times n}$ task in which Alice and Bob must complete $\textbf{B}_{84}$ $n$ times in parallel, one can argue for a lower bound on the mutual information shared by Alice and Bob which scales linearly with $n$; in the holographic setting, one can take $n$ to scale with any function smaller than $O(1/G)$ without incurring backreaction, implying in particular that the mutual information is larger than $O(1)$. 
For additional details, see \cite{May:2019odp}. 

\subsubsection*{The converse direction}

One might already wonder at this stage whether a converse statement to theorem \ref{thm:CWT}, i.e. a statement that a connected wedge $\mathcal{E}(\V_{1} \cup \V_{2})$ is sufficient for bulk scattering, could possibly hold. In fact, \cite{May:2019odp} provides a simple counter-example to this possibility. In this counter-example, one begins by considering vacuum AdS, with boundary regions $\V_{1}, \V_{2}$ and $\X_{1}, \X_{2}$ each an interval of size $\pi/2$ in the $t=0$ slice. In this situation, $\mathcal{S}$ is marginally non-empty, consisting of a single point. 
One can then modify the state by introducing a spherically symmetric dust ball in the centre of AdS, which results in a finite time delay for null geodesics and renders $\mathcal{S}$ empty. A small increase in the angular size of $\V_{1}, \V_{2}$ can then result in a connected entanglement wedge while $\mathcal{S} = \emptyset$ persists.

\subsubsection*{A lower bound from the ridge}

Beyond establishing $I(\V_{1} : \V_{2}) = O(1/G)$, the proof technique of \cite{May:2019odp} in fact yields an explicit geometrical lower bound on the mutual information. In particular, one can argue in the case of bulk-only scattering configurations that the bulk codimension-2 surface
\begin{equation}
    \mathfrak{r} \equiv \partial \mathcal{J}^{+}[\mathcal{E}[\V_{1}]] \cap \partial \mathcal{J}^{+}[\mathcal{E}[\V_{2}]] \cap \mathcal{J}^{-}[r_{1}] \cap \mathcal{J}^{-}[r_{2}] \: ,
\end{equation}
referred to as the \textit{ridge}, is non-empty, and its area lower bounds the mutual information by
\begin{equation} \label{eq:ridgelb}
    I(\V_{1} : \V_{2}) \geq 2 \times \frac{\text{Area}[\mathfrak{r}]}{4G} \: .
\end{equation}

\subsubsection*{Boundary points versus boundary regions}

An improved version of the connected wedge theorem, which subsumes that stated above, replaces the input and output points $c_{1}, c_{2}, r_{1}, r_{2}$ with boundary regions $\hat{\mathcal{C}}_{1}, \hat{\mathcal{C}}_{2}, \hat{\mathcal{R}}_{1}, \hat{\mathcal{R}}_{2}$. The decision regions and boundary scattering region are defined analogously to before, while the bulk scattering region is now defined by
\begin{equation}
    \mathcal{S} \equiv \mathcal{J}^{+} [ \mathcal{E}[\hat{\mathcal{C}}_{1}]] \cap \mathcal{J}^{+} [ \mathcal{E}[\hat{\mathcal{C}}_{2}]] \cap \mathcal{J}^{-} [ \mathcal{E}[\hat{\mathcal{R}}_{1}]] \cap \mathcal{J}^{-} [ \mathcal{E}[\hat{\mathcal{R}}_{2}]] \: .
\end{equation}
Subject to the assumption $\hat{\mathcal{C}}_{i} \subseteq \hat{\mathcal{V}}_{i}$, a connected wedge theorem can again be proven.

\subsection{The generalized connected wedge proposal}

It was already observed in \cite{May:2019odp} that the scattering region $\mathcal{S}$ is in fact contained in the entanglement wedge of $\V_{1} \cup \V_{2}$,
\begin{equation}
    \mathcal{S} \subseteq \mathcal{E}[\V_{1} \cup \V_{2}] \: ,
\end{equation}
and likewise (by time reversal) in the entanglement wedge of $\W_{1} \cup \W_{2}$. 
It follows that the spacetime region
\begin{equation}
    \mathcal{S}_{E} = \mathcal{E}[\V_{1} \cup \V_{2}] \cap \mathcal{E}[\W_{1} \cup \W_{2}] \: ,
\end{equation}
is guaranteed to contain the scattering region $\mathcal{S}$, so that a putative sufficient condition for $\mathcal{S}_{E} \neq \emptyset$ will be weaker than a sufficient condition for $\mathcal{S} \neq \emptyset$. This can be seen as a motivation for the GCW proposal of Leutheusser and Liu, which we state below.

\begin{proposal} \label{prop:LL}
(Leutheusser-Liu \cite{Leutheusser:2024yvf}) Let $c_{1}, c_{2}, r_{1}, r_{2}$ be such that $\hat{\mathcal{S}} = \emptyset$. Then $\mathcal{E}[\V_{1} \cup \V_{2}]$ is connected if and only if $\mathcal{S}_{E} \neq \emptyset$.
\end{proposal}

\noindent
In more detail, the stated assumptions in the proposal are as follows:
\begin{itemize}
    \item The null energy condition holds in the bulk spacetime.
    \item There exist bulk Cauchy surfaces $\Sigma_{V}, \Sigma_{W}$ to the past of $\mathcal{E}[\V_{1} \cup \V_{2}]$ and the future of $\mathcal{E}[\W_{1} \cup \W_{2}]$ respectively such that the region between $\Sigma_{V}$ and $\Sigma_{W}$ does not contain singularities. 
\end{itemize}
Moreover, it is assumed that the bulk spacetime should correspond to a pure state in the CFT. We note that the ``no singularities'' condition would be implied by AdS-hyperbolicity. 

In what follows we will assume that we are in a pure state. This lets us avoid horizons complicating the focusing arguments (see eg.~\cite{May:2022clu}), and allows us to implicitly assume that $\gamma_{\hat{\mathcal{A}}}=\gamma_{\hat{\mathcal{A}}'}$ in various steps.

\section{Implications of connected wedges} \label{sec:proof}

In this section, we will see that a certain reformulation of the GCW proposal relating connectedness of $\mathcal{E}[\V_{1} \cup \V_{2}]$, $\mathcal{E}[\W_{1} \cup \W_{2}]$ to the emptiness of $\mathcal{S}_{E}$ can be proven in generality.

\subsection{Connected wedges imply non-empty \texorpdfstring{$\mathcal{S}_{E}$}{}}\label{sec:main}

As we will show in section \ref{sec:counter-ex}, one can identify situations in which $\mathcal{E}[\V_{1} \cup \V_{2}]$ is connected but $\mathcal{S}_{E} = \emptyset$, seemingly contravening the GCW proposal \ref{prop:LL}. We will see for example that it is possible to obtain holographic states for which $\mathcal{E}[\V_{1} \cup \V_{2}]$ is connected but $\mathcal{E}[\W_{1} \cup \W_{2}]$ is disconnected, which immediately puts the statement in proposal \ref{prop:LL} in conflict with its time reversal. 
However, we nonetheless have the following proposition (see section \ref{sec:review} for definitions).
\begin{proposition}\label{GCWT2}
    Let $c_{1}, c_{2}, r_{1}, r_{2}$ be such that $\V_{1}, \V_{2}, \W_{1}, \W_{2}$ are all non-empty and non-intersecting (so that there is no boundary scattering). Suppose that $\mathcal{E}[\V_{1} \cup \V_{2}]$ and $\mathcal{E}[\W_{1} \cup \W_{2}]$ are both connected. Then $\mathcal{S}_{E} \neq \emptyset$. 
\end{proposition}
The proof of this proposition is closely related to that of May, Penington, and Sorce \cite{May:2019odp} in proving theorem \ref{thm:CWT}, and the assumptions will be the same as in that case (see section \ref{sec:review}).
A sketch of the proof is as follows.
\begin{proof}
We proceed by contradiction, supposing that $\mathcal{E}[\V_{1} \cup \V_{2}]$ and $\mathcal{E}[\W_{1} \cup \W_{2}]$ are both connected, but $\mathcal{S}_{E} = \emptyset$. 
It will be convenient here and in subsequent sections to introduce the notation
\begin{equation}
    \rho_{\X}^{\pm} \equiv \partial \mathcal{J}^{\pm} [ \mathcal{E}[\X_{1}]] \cap \partial \mathcal{J}^{\pm} [ \mathcal{E}[\X_{2}]] \: , \qquad \rho_{\Y}^{\pm} \equiv \partial \mathcal{J}^{\pm} [ \mathcal{E}[\Y_{1}]] \cap \partial \mathcal{J}^{\pm} [ \mathcal{E}[\Y_{2}]] \: ;
\end{equation}
these are codimension-2 surfaces resulting from the intersection of future/past-directed ingoing lightsheets from the RT surfaces from the $\X_{i}$ and $\Y_{i}$ regions respectively, which we refer to as \textit{extended ridges}. Let us record a couple of basic facts about these extended ridges:
\begin{itemize}
    \item They are connected, codimension-2 surfaces (i.e. with the topology of a line) anchored at the AdS boundary. 
    \item For connected $\mathcal{E}[\V_{1} \cup \V_{2}]$ and $\mathcal{E}[\W_{1} \cup \W_{2}]$, one has containment $\rho_{\X}^{\pm} \subset \mathcal{E}[\V_{1} \cup \V_{2}]$ and $\rho_{\Y}^{\pm} \subset \mathcal{E}[\W_{1} \cup \W_{2}]$.
    \item For connected and non-intersecting $\mathcal{E}[\V_{1} \cup \V_{2}]$ and $\mathcal{E}[\W_{1} \cup \W_{2}]$, there is no point in $\rho_{\X}^{+}$ which lies to the future of a point in $\rho_{\Y}^{+}$. 
\end{itemize}
The first comment follows from AdS-hyperbolicity, and the second is clear from the definitions. 
The intuitive reason for the third is that the alternative would in turn require all of $\mathcal{E}[\W_{1} \cup \W_{2}]$ to ``pass under'' $\mathcal{E}[\V_{1} \cup \V_{2}]$ instead of ``passing over'' it, which cannot occur; we explain this more precisely in appendix \ref{app:vw} for completeness.

\begin{figure}
    \centering   
    \includegraphics[height=5cm]{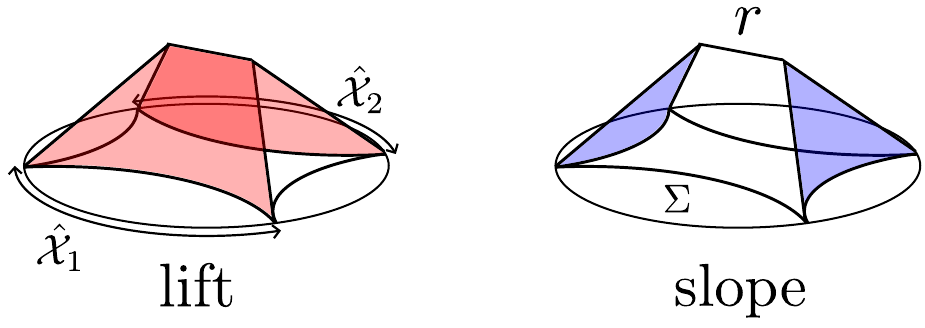}
    \caption{Lift $\lambda$ (red) and slope $\sigma$ (blue), defined in \eqref{eq:lift} and \eqref{eq:slope} and constituting the null membrane. We also indicate the boundary regions $\X_{1}, \X_{2}$, the slice $\Sigma$, and the ridge $r$. }\label{fig:nullmem}
\end{figure}

Now, let $\Sigma$ be some complete achronal slice containing the RT surface for $\V_{1} \cup \V_{2}$. 
We will be interested in constructing a codimension-1 \textit{null membrane} $\mathcal{N}$, which consists of a \textit{lift} surface
\begin{equation} \label{eq:lift}
    \lambda = \partial \mathcal{J}^{+} [ \mathcal{E}[\X_{1} \cup \X_{2}] ] \cap \mathcal{J}^{-}[\mathcal{E}[\Y_{1}]] \cap \mathcal{J}^{-}[\mathcal{E}[\Y_{2}]] \: ,
\end{equation}
and a \textit{slope} surface
\begin{equation} \label{eq:slope}
    \sigma = \partial \left[ \mathcal{J}^{-}[\mathcal{E}[\Y_{1}]]  \cap \mathcal{J}^{-}[\mathcal{E}[\Y_{2}]] \right] \cap \mathcal{J}^{-} \left[ \partial \mathcal{J}^{+} \left[ \mathcal{E}[\X_{1} \cup \X_{2}] \right] \right] \cap \mathcal{J}^{+}[\Sigma] \: .
\end{equation}
See figure \ref{fig:nullmem} for an illustration. The lift $\lambda$ consists of parts of the ingoing forward lightsheets from the RT surfaces for $\X_{1}$ and $\X_{2}$, up to a codimension-2 surface, the \textit{ridge} $r$, where they collide. The ridge is just the subset of the extended ridge $\rho_{\X}^{+}$ in the past of the entanglement wedges of the $\Y$ regions,
\begin{equation}
    r = \rho_{\X}^{+} \cap \mathcal{J}^{-}[\mathcal{E}[\Y_{1}]] \cap \mathcal{J}^{-}[\mathcal{E}[\Y_{2}]] \: ,
\end{equation}
so in particular it is not boundary-anchored.
The slope $\sigma$ then consists of parts of the ingoing backward lightsheets from the RT surfaces for $\Y_{1}$ and $\Y_{2}$, lying in the past of the lift, but in the future of $\Sigma$. 

\begin{figure}
    \centering   
    \includegraphics[height=6cm]{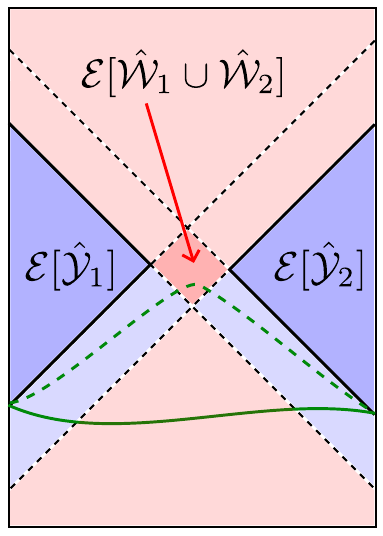}
    \caption{Cross-section of spacetime highlighting its decomposition into the two wedges in \eqref{eq:endpoint_wedges} (blue) and the two wedges in \eqref{eq:nonendpoint_wedges} (red). The darker shaded blue regions correspond to $\mathcal{E}[\Y_{1}]$ and $\mathcal{E}[\Y_{2}]$, while the darker shaded red region corresponds to $\mathcal{E}[\W_{1} \cup \W_{2}]$. We argue that the trajectory of the extended ridge $\rho_{\X}^{+}$ must correspond to the solid green line rather than the dashed green line, since $\mathcal{S}_{E} = \emptyset$.}\label{fig:ridgewedges}
\end{figure}

The key observation now is that, since $\mathcal{S}_{E} = \emptyset$ by assumption, we can show that the spacelike ridge $r$ is non-empty, and the topology of $\mathcal{N}$ is as shown in figure \ref{fig:nullmem}. 
Note that the endpoints of $\rho_{\X}^{+}$ lie within 
\begin{equation} \label{eq:endpoint_wedges}
    \mathcal{J}^{-}[\mathcal{E}[\Y_{1}]] \cap \mathcal{J}^{-}[\mathcal{E}[\Y_{2}]]^{c} \qquad \text{and} \qquad \mathcal{J}^{-}[\mathcal{E}[\Y_{1}]]^{c} \cap \mathcal{J}^{-}[\mathcal{E}[\Y_{2}]]
\end{equation}
respectively. Since the spacetime splits into four regions, i.e. the above two regions and 
\begin{equation} \label{eq:nonendpoint_wedges}
    \mathcal{J}^{-}[\mathcal{E}[\Y_{1}]] \cap \mathcal{J}^{-}[\mathcal{E}[\Y_{2}]] \: , \qquad \mathcal{J}^{-}[\mathcal{E}[\Y_{1}]]^{c} \cap \mathcal{J}^{-}[\mathcal{E}[\Y_{2}]]^{c} \: ,
\end{equation}
connectedness of $\rho_{\X}^{+}$ implies that it must pass through one of these latter two regions (or both);\footnote{It could also pass directly from $\mathcal{J}^{-}[\mathcal{E}[\Y_{1}]] \cap \mathcal{J}^{-}[\mathcal{E}[\Y_{2}]]^{c}$ to $\mathcal{J}^{-}[\mathcal{E}[\Y_{1}]]^{c} \cap \mathcal{J}^{-}[\mathcal{E}[\Y_{2}]]$ through the edge $\partial \mathcal{J}^{-}[\mathcal{E}[\Y_{1}]] \cap \partial \mathcal{J}^{-}[\mathcal{E}[\Y_{2}]]$; in that case, the ridge would have at least one point, and that point would be in $\mathcal{S}_{E}$, contradicting that $\mathcal{S}_{E}$ is empty.} see figure \ref{fig:ridgewedges}.
We observe that the closure of $\mathcal{J}^{-}[\mathcal{E}[\Y_{1}]]^{c} \cap \mathcal{J}^{-}[\mathcal{E}[\Y_{2}]]^{c}$ is equivalent to $\mathcal{J}^{+}[ \mathcal{E}[\W_{1} \cup \W_{2}] ]$, and that, were $\rho_{\X}^{+}$ to pass through this region, connectedness would require it to pass through either $\mathcal{E}[\W_{1} \cup \W_{2}]$ or $\mathcal{J}^{+}[\rho_{\Y}^{+}]$ (or both).\footnote{Again, $\mathcal{S}_{E}$ would not be empty in the edge case.} Since $\mathcal{S}_{E} = \emptyset$, neither of these options is possible (using the second and third facts about extended ridges mentioned earlier). 
It follows that $\rho_{\X}^{+}$ must pass through $\mathcal{J}^{-}[\mathcal{E}[\Y_{1}]] \cap \mathcal{J}^{-}[\mathcal{E}[\Y_{2}]]$, and thus $r$ is non-empty. One can follow the logic of \cite{May:2019odp}, focusing along the null membrane (``up the ridge'' and ``down the slope'') to construct a ``contradiction surface'' on $\Sigma$ which is homologous to $\V_{1} \cup \V_{2}$ but has smaller area than the putative RT surface, contradicting maximin; we omit the details, since this part of the argument is precisely as in \cite{May:2019odp}. The proposition follows. 
\end{proof}

\subsection{An upper bound on the mutual information}\label{sec:mi_bound}

In this section we will construct an analogue $\mathfrak{r}_{E}$ of the ``ridge'' $\mathfrak{r}$ in~\cite{May:2019odp} for our scattering region ${\cal S}_E$ (not to be confused with the ridge $r$ introduced in the previous subsection), and show that its area gives an \textit{upper} bound for the mutual information when both wedges are connected.\footnote{We thank Athira Arayath for extensive discussions on this proof and understanding edge cases for the topology of the ${\cal E}[\V_1\cup \V_2] \cap {\cal E}[\W_1\cup \W_2]$ intersection. This upper bound will be elaborated upon and generalized to a statement about scattering region entropies in~\cite{ap}.} 
This will give a somewhat more quantitative characterization of the size of $\mathcal{S}_{E}$ when the $\V_{i}$ and $\W_{i}$ each have $O(1/G)$ mutual information: not only is $\mathcal{S}_{E}$ non-empty in this case, but a measure of its size in terms of the area of the ridge $\mathfrak{r}_{E}$ can be lower bounded by the mutual information between either the $\V_{i}$ or the $\W_{i}$. 

\begin{figure}
    \centering   
    \includegraphics[height=8cm]{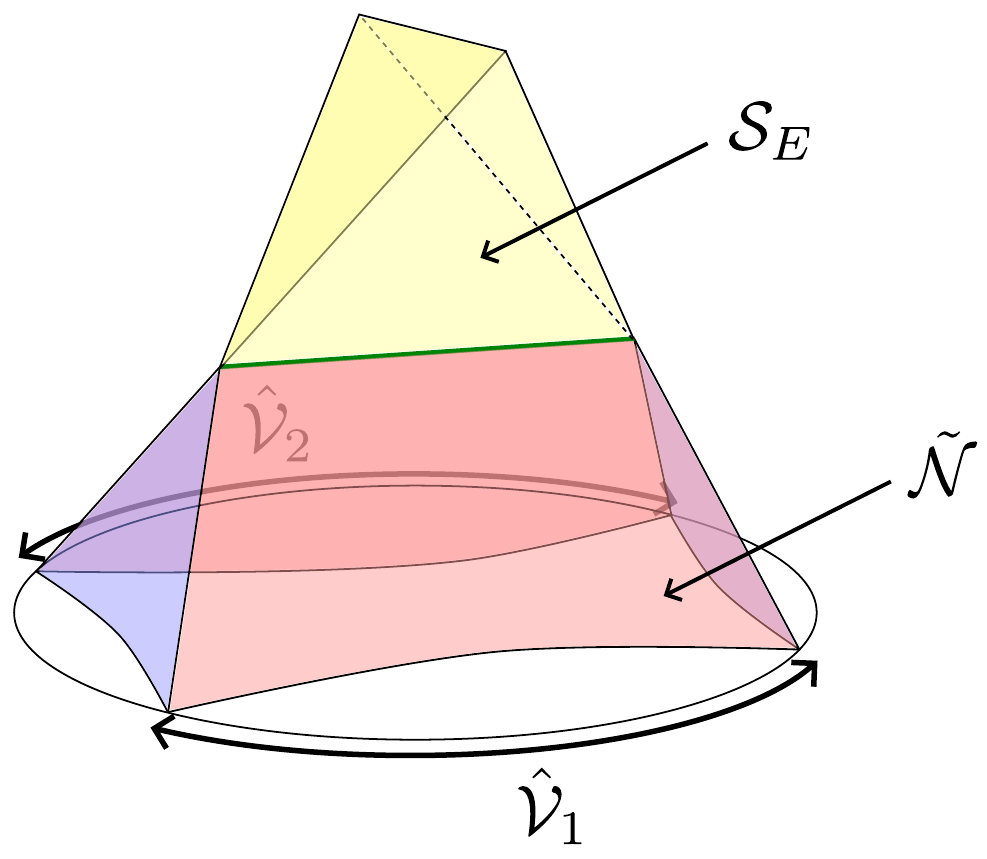}
    \caption{Depiction of the null membrane $\tilde{\mathcal{N}}$ and scattering region $\mathcal{S}_{E}$ for the case that $\rho_{\Y}^{-} \cap \mathcal{J}^{-}[\Sigma] = \emptyset$. The lift $\tilde{\lambda}$ and slope $\tilde{\sigma}$ are depicted in red and blue respectively, while the scattering region $\mathcal{S}_{E}$ is illustrated as a yellow tetrahedron, and the ridge $\mathfrak{r}_{E}$ is highlighted in green. Focusing upward along the slope and downward along the lift results in the upper bound \eqref{eq:lbmi}. }\label{fig:ubfig}
\end{figure}

Let $\Sigma$ denote a maximin slice for the boundary region $\V_{1} \cup \V_{2}$. 
The definition of the ridge $\mathfrak{r}_{E}$ in this context is again given in terms of the extended ridges $\rho_{\X}^{\pm}, \rho_{\Y}^{\pm}$ defined in the previous subsection. 
Throughout this subsection, we will consider the case that 
\begin{equation} \label{eq:ridge_assumption}
    \text{assumption:} \quad \rho_{\Y}^{-} \cap \mathcal{J}^{-}[\Sigma] = \emptyset \: ,
\end{equation}
meaning that the extended ridge $\rho_{\Y}^{-}$ remains entirely to the future of $\Sigma$; the more general setting, for which the structure of the ridge becomes more complicated and in particular depends on the choice of $\Sigma$, will be discussed in \cite{ap}, but we prefer to omit those details here to illustrate the basic point. In this case, we will again define a codimension-1 null membrane $\tilde{\mathcal{N}}$, this time consisting of a lift surface
\begin{multline}
    \tilde{\lambda} = \left( \left( \partial \mathcal{J}^{-}[\mathcal{E}[\Y_{1}]] \cap \mathcal{J}^{-}[\mathcal{E}[\Y_{2}]] \right) \cup \left( \partial \mathcal{J}^{-}[\mathcal{E}[\Y_{2}]] \cap \mathcal{J}^{-}[\mathcal{E}[\Y_{1}]] \right) \right) \\
    \cap \mathcal{J}^{+}[\Sigma] \setminus \mathcal{J}^{+}[\mathcal{E}[\X_{1} \cup \X_{2}]] 
\end{multline}
and a slope surface
\begin{equation}
    \tilde{\sigma} = \partial \mathcal{J}^{+}[\mathcal{E}[\X_{1} \cup \X_{2}]] \cap \mathcal{J}^{-}[\mathcal{E}[\Y_{1}]] \cap \mathcal{J}^{-}[\mathcal{E}[\Y_{2}]] \: .
\end{equation}
See figure \ref{fig:ubfig}. 
Unlike the null membrane in the previous subsection and that appearing in \cite{May:2019odp}, the lift is now part of the boundary of the \textit{past} of some regions, while the slope is part of the boundary of the \textit{future} of a region; this reverses the direction of focusing as compared to those settings, which is what will lead to an upper bound rather than a lower bound on $I(\V_{1} : \V_{2})$. We can then define the ridge by
\begin{equation}
    \mathfrak{r}_{E} \equiv \rho_{\Y}^{-} \cap \mathcal{E}[\V_{1} \cup \V_{2}] \: ,
\end{equation}
as shown in figure \ref{fig:ubfig}. This quantity will be non-empty whenever $\mathcal{S}_{E}$ is non-empty, at least in the present setting with connected wedges and the assumption \eqref{eq:ridge_assumption}. 

With the above definitions, we now claim that
\begin{equation} \label{eq:lbmi}
    I(\V_{1} : \V_{2}) \leq 2 \times \frac{\text{Area}[\mathfrak{r}_{E}]}{4 G} \: .
\end{equation}
This can be compared to the analogous \textit{lower} bound on the mutual information from the area of the ordinary ridge in \eqref{eq:ridgelb}. 
As presaged above, this result follows from focusing upward along the slope from the RT surfaces for $\X_{1}$ and $\X_{2}$ to the ``cusps''
\begin{equation}
    \mathfrak{c}_{ij} \equiv \partial \mathcal{J}^{+}[\mathcal{E}[\X_{i}]] \cap \partial \mathcal{J}^{-}[\mathcal{E}[\Y_{j}]] \: ,
\end{equation}
then downward along the lift from the cusps and the ridge to the surface
\begin{equation}
    \mathfrak{s} \equiv \tilde{\lambda} \cap \Sigma \: .
\end{equation}
This process gives us inequalities
\begin{equation}
    \begin{split}
        & \text{Area}[\gamma[\X_{1}]] + \text{Area}[\gamma[\X_{2}]] \geq \sum_{i, j} \text{Area}[\mathfrak{c}_{ij}] \\
        & \sum_{i, j} \text{Area}[\mathfrak{c}_{ij}] + \text{Area}[\mathfrak{r}_{E}] \geq \text{Area}[\mathfrak{s}] \: .
    \end{split}
\end{equation}
Combining these inequalities, recalling that $\gamma[\V_{1} \cup \V_{2}] = \gamma[\X_{1}] \cup \gamma[\X_{2}]$, and using the maximin property to infer
\begin{equation}
    \text{Area}[\mathfrak{s}] \geq \text{Area}[\gamma[\V_{1}]] + \text{Area}[\gamma[\V_{2}]] \: ,
\end{equation}
we obtain
\begin{equation}
    \text{Area}[\mathfrak{r}_{E}] \geq \text{Area}[\gamma[\V_{1}]] + \text{Area}[\gamma[\V_{2}]] - \text{Area}[\gamma[\V_{1} \cup \V_{2}]] \: ,
\end{equation}
giving the desired result \eqref{eq:lbmi}.

\subsection{A GCW theorem with converse}\label{sec:converse}

We will now show that the bulk region 
\begin{equation}
    \tilde{\mathcal{S}}_{E} \equiv \left( \mathcal{E}[\V_{1} \cup \V_{2}] \setminus (\mathcal{E}[\V_{1}] \cup \mathcal{E}[\V_{2}]) \right) \cap \left( \mathcal{E}[\W_{1} \cup \W_{2}] \setminus (\mathcal{E}[\W_{1}] \cup \mathcal{E}[\W_{2}]) \right)
\end{equation}
defined in the introduction is non-empty if and only if both wedges are connected. That one or both wedges being disconnected implies empty $\tilde{\mathcal{S}}_{E}$ is immediate from the definition. 
For the other direction, what remains to be shown is that connected wedges for $\V_{1} \cup \V_{2}$ and $\W_{1} \cup \W_{2}$ imply not only that $S_{E}$ is non-empty, but that it includes a point outside of $\mathcal{E}[\V_{1}]\cup \mathcal{E}[\V_{2}]\cup \mathcal{E}[\W_{1}]\cup \mathcal{E}[\W_{2}]$.

However, provided that $\mathcal{E}[\V_{1}] \cup \mathcal{E}[\V_{2}] \subset \text{int}[\mathcal{E}[\V_{1} \cup \V_{2}]]$ and likewise for regions $\W_{1}, \W_{2}$, this is straightforward to establish. Indeed, in this case, if $\mathcal{S}_{E}$ only consisted of points in $\mathcal{E}[\V_{1}]\cup \mathcal{E}[\V_{2}]\cup \mathcal{E}[\W_{1}]\cup \mathcal{E}[\W_{2}]$, then one of $\mathcal{E}[\V_{i}]$ and $\mathcal{E}[\W_{j}]$ would have non-empty overlap; without loss of generality we assume $\mathcal{E}[\V_{1}] \cap \mathcal{E}[\W_{1}] \neq \emptyset$. In particular, if $\mathcal{E}[\V_{1}]$ and $\mathcal{E}[\W_{1}]$ are each connected regions (as they should be in the classical regime), there must be a point $p \in \partial \mathcal{E}[\V_{1}] \cap \partial \mathcal{E}[\W_{1}]$. But if $\mathcal{E}[\V_{1}] \subset \text{int}[\mathcal{E}[\V_{1} \cup \V_{2}]]$ and $\mathcal{E}[\W_{1}] \subset \text{int}[\mathcal{E}[\W_{1} \cup \W_{2}]]$, then there is an open neighbourhood of $p$ contained in $\mathcal{E}[\V_{1} \cup \V_{2}] \cap \mathcal{E}[\W_{1} \cup \W_{2}]$. Since $p$ is on the boundary of $\mathcal{E}[\V_{1}]$ and $\mathcal{E}[\W_{1}]$, this open neighbourhood will contain points outside of both; such points would be in $\tilde{\mathcal{S}}_{E}$, as desired. 

The only assumption to be justified in the above argument is that one has containment $\mathcal{E}[\V_{1}] \cup \mathcal{E}[\V_{2}] \subset \text{int}[\mathcal{E}[\V_{1} \cup \V_{2}]]$ whenever the wedge is connected, and likewise for $\W_{1}, \W_{2}$. This should in fact be immediate from the properties of the  
future/past of a set. Specifically, we expect that for regions $A, B$ on spatial slice $\Sigma$ satisfying $A \subset \text{int}[B]$, one has $\mathcal{J}^{\pm}[A] \subset \text{int}[\mathcal{J}^{\pm}[B]]$. Applying this with $A$ taken to be the union of the entanglement wedges of $\X_{1}, \X_{2}$ and $B$ taken to be the entanglement wedge of $\V_{i}'$, we find that $\partial \mathcal{J}^{\pm}[\mathcal{E}[\X_{1}] \cup \mathcal{E}[\X_{2}]]$ and $\partial \mathcal{J}^{\pm}[\mathcal{E}[\V_{i}']]$ can't intersect, except potentially at the conformal boundary; since these quantities make up $\partial \mathcal{E}[\V_{1} \cup \V_{2}]$ and $\partial \mathcal{E}[\V_{i}]$ respectively, this is the desired conclusion.

\section{No other logical relationships} \label{sec:counter-ex}

Now let us see if we can make the statement from section \ref{sec:proof} stronger. Given the results of that section, the claim closest to the proposal in~\cite{Leutheusser:2024yvf}
that we might hope to prove is
\be\label{prop2}
\mathcal{E}[\V_1\cup \V_2]~{\rm and}~\mathcal{E}[\W_1\cup \W_2] {\rm ~connected~} \overset{?}{\Leftrightarrow} {\cal S}_E\neq \emptyset \: .
\ee
More generally, there are eight possible logical relationships between the connectedness or disconnectedness of the entanglement wedges and emptiness or non-emptiness of $\mathcal{S}_{E}$. Indeed, we can consider a statement of the form $A \implies B$, where $A$ is taken to be one of the four possiblities
\begin{enumerate}
    \item $\mathcal{E}[\V_{1} \cup \V_{2}]$ disconnected, $\mathcal{E}[\W_{1} \cup \W_{2}]$ disconnected,
    \item $\mathcal{E}[\V_{1} \cup \V_{2}]$ disconnected, $\mathcal{E}[\W_{1} \cup \W_{2}]$ connected,
    \item $\mathcal{E}[\V_{1} \cup \V_{2}]$ connected, $\mathcal{E}[\W_{1} \cup \W_{2}]$ disconnected,
    \item $\mathcal{E}[\V_{1} \cup \V_{2}]$ connected, $\mathcal{E}[\W_{1} \cup \W_{2}]$ connected,
\end{enumerate}
and $B$ is taken to be one of the two possibilities $\mathcal{S}_{E} = \emptyset$ or $\mathcal{S}_{E} \neq \emptyset$.\footnote{Of course, we can switch from a situation with $\mathcal{E}[\V_{1} \cup \V_{2}]$ connected and $\mathcal{E}[\W_{1} \cup \W_{2}]$ to the opposite via time reversal.} The purpose of this section will be to show counter-examples for all of these possible logical relationships, except for the implication 
\be\label{eq:sec3summ}
\mathcal{E}[\V_1\cup \V_2]~{\rm and}~\mathcal{E}[\W_1\cup \W_2] {\rm ~connected~} \implies {\cal S}_E\neq \emptyset 
\ee
shown in the previous section.

\subsection{Scattering in the conical defect}\label{sec:def}

As mentioned in section \ref{sec:review}, the GCW proposal holds in vacuum AdS$_{3}$. While it is straightforward to show this via geometrical calculations in the bulk, it can also be argued for directly on the basis of the CFT: the boundaries of regions $\V_{1}, \V_{2}$ and $\W_{1}, \W_{2}$ correspond to the same values of the conformal cross ratios, so the mutual information, computed via a vacuum correlation function of twist operators, must be equal. 

Perhaps the next simplest class of states to consider from the bulk perspective are those  corresponding to a single conical defect in AdS$_{3}$. The metric of the asymptotically AdS$_{3}$ conical defect spacetime, setting $\lads=1$, is
\begin{align}
     \dd{s}^2= -(r^2 - M) \dd{t}^2 + \frac{\dd{r}^2}{r^2-M}+r^2 \dd{\phi}^2,
\end{align}
where $-1<M<0$ parametrizes the mass of the defect, and we have coordinate range
\begin{equation}
    r\geq 0 \: , \quad 0 \leq \phi < 2\pi \: , \quad t \in \mathbb{R} \: .
\end{equation}
Notice that the same metric yields the BTZ black hole metric when $M\geq 0$ and that for $M=-1$ it reduces to pure AdS$_3$.
We will see that, despite its simplicity, this class of spacetimes is rich enough to provide counter-examples for all possible logical relationships between the connectedness of entanglement wedges $\mathcal{E}[\V_{1} \cup \V_{2}]$, $\mathcal{E}[\W_{1} \cup \W_{2}]$ and the emptiness or non-emptiness of $\mathcal{S}_{E}$, except for the statement \eqref{eq:sec3summ} shown in section \ref{sec:proof}.\footnote{Holographic scattering and the connected wedge theorem have previously been studied in this spacetime for various choices of kinematics $\V_{i}, \W_{i}$ \cite{Caminiti:2024ctd}. The kinematics we consider here are more general, so many of the counter-examples that we will present here are not amongst the cases considered in that work.}

\begin{figure}
    \centering
    \begin{tikzpicture}[scale=0.8]
\draw[thick] (-8, 2.5) rectangle (8, -7.5);

\draw[->, thick] (-8.5, 1.5) -- (-8.5, 2) node[above] {$t $};
\draw[->, thick] (7,-8) -- (7.5,-8) node[right] {$\phi$};

\draw[fill=red!40, thick]
  (-8, 0) -- (-7,1) -- (0.5, -6.5) -- (-0.5, -7.5) -- cycle;
  \node[red] at (-3.75, -3.25) {$\hat{\mathcal{V}}_1$};
  
\draw[fill=red!40, thick]
  (1.4, -6.4) -- (2.1, -5.7) -- (1.8, -5.4)-- (1.1, -6.1) -- cycle;
  \node[red] at (2, -6.3) {$\hat{\mathcal{V}}_2$};

\draw[fill=blue!40, thick]
  (-6.5, 1.5) -- (-5.8, 2.2) -- (1.7, -5.3) -- (1, -6) -- cycle;
\node[blue] at (-2.4, -1.9) {$\hat{\mathcal{W}}_1$};

\draw[fill=blue!40, thick]
  (-8, 0.2) -- (-7.1, 1.1) -- (-7.4, 1.4) -- (-8, 0.8) -- cycle;
\draw[fill=blue!40, thick]
  (8, 0.2) --(7.9, 0.1)-- (7.6, 0.4) -- (8, 0.8) -- cycle;
\node[blue] at (-7.6, 1.8) {$\hat{\mathcal{W}}_2$};

\draw[densely dashed, red, thick] (-0.5, -7.5) -- (7.5, 0.5);
\draw[densely dashed, red, thick] (8, 0) -- (7.5, 0.5);
\draw[densely dashed, red, thick] (-0.5, -7.5) -- (-8,0);

\draw[densely dashed, red, thick] (1.4, -6.4) -- (8, 0.2);
\draw[densely dashed, red, thick] (-8, 0.2) -- (-6.6, 1.6);
\draw[densely dashed, red, thick] (1.4, -6.4) -- (-6.6, 1.6);

\draw[densely dashed, blue, thick] (-5.8, 2.2) -- (2.2, -5.8);
\draw[densely dashed, blue, thick] (-5.8, 2.2) -- (-8,0);
\draw[densely dashed, blue, thick] (8, 0) -- (2.2, -5.8);

\draw[densely dashed, blue, thick] (8, 0.8) -- (0.6, -6.6);
\draw[densely dashed, blue, thick] (-7.4, 1.4) -- (0.6, -6.6);
\draw[densely dashed, blue, thick] (-7.4, 1.4) -- (-8, 0.8);


\draw[fill=black] (-8, 0) circle (2pt);
\draw[fill=black] (0.5, -6.5) circle (2pt);

\draw[fill=black] (1.1, -6.1) circle (2pt) node[below] {\scriptsize $E$};
\draw[fill=black] (2.1, -5.7) circle (2pt) node[right] {\scriptsize $F$};

\draw[fill=black] (-6.5, 1.5) circle (2pt) node[below] {\scriptsize $A$};
\draw[fill=black] (1.7, -5.3) circle (2pt) node[above] {\scriptsize $B$};

\draw[fill=black] (7.6, 0.4) circle (2pt) node[below] {\scriptsize $C$};
\draw[fill=black] (-7.1, 1.1) circle (2pt) node[above] {\scriptsize $D$};

\end{tikzpicture}
\caption{Boundary set-up for scattering problem counter-examples. The dashed blue lines represent the past of $r_1$ (for $\hat{\mathcal{W}}_1$) and $r_2$ (for $\hat{\mathcal{W}}_2$), while the dashed red lines represent the future of $c_1$ (for $\hat{\mathcal{V}}_1$) and $c_2$ (for $\hat{\mathcal{V}}_2$). We label the left and right-most points of $\hat{\mathcal{W}}_1$, $\hat{\mathcal{W}}_2$ and $\hat{\mathcal{V}}_2$, as they will be used in our analysis. The region $\hat{\mathcal{V}}_1$ will be kept fixed. Notice that we are interested in the regime in which $\hat{\mathcal{W}}_1$ has angular length strictly larger than $\pi$.}
\label{fig:setupcounterexample}
\end{figure}
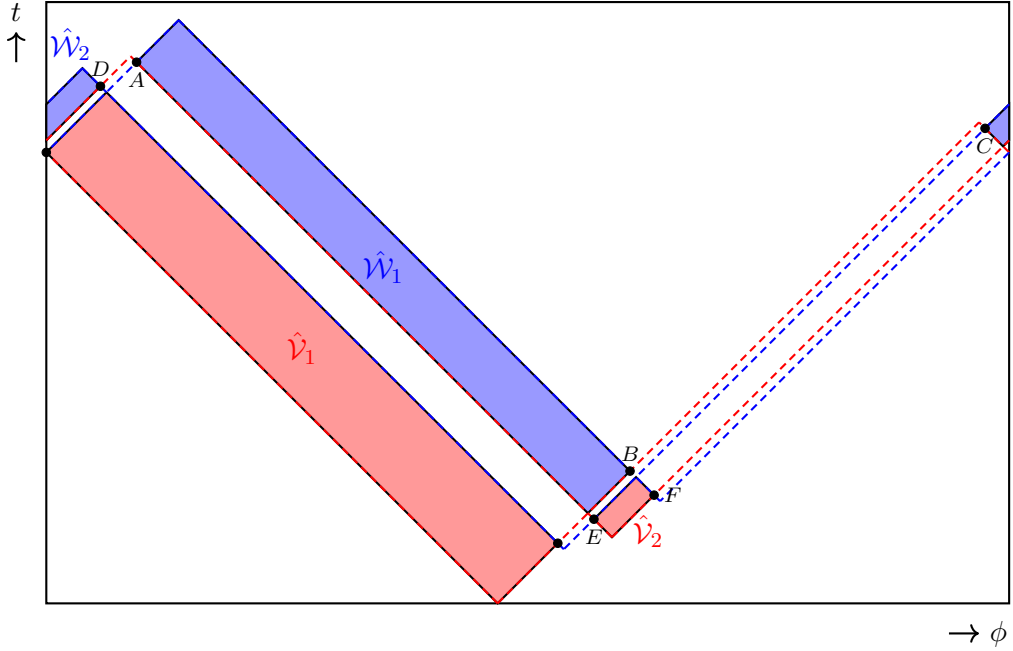

For what follows we will want to set up a class of configurations where we can tune between the $\mathcal{E}[\V_1\cup \V_2]$ and $\mathcal{E}[\W_1\cup \W_2]$ being connected or disconnected. Consider figure~\ref{fig:setupcounterexample}.
We parametrize the boundary points by $(t, \phi)$. Without loss of generality, we set the leftmost point of $\hat{\mathcal{V}}_1$ to $(0, 0)$. In principle, its rightmost point can be left free, but for the sake of the examples here, we will set it to $(-\frac{13 \pi}{16}, \frac{17 \pi}{16})$.\footnote{Importantly, for the counter-examples we consider, it turns out that we require that the coordinate range $\Delta \phi$ spanned by $\V_{1}$ must be greater than $\pi$, so in particular $\mathcal{C}[\V_{1}] \subsetneq \mathcal{E}[\V_{1}]$.} Given this choice, the other points are given by
\begin{align}\label{eq:phi}
    A&= (\phi_A, \phi_A) \nonumber\\
    B&= (\phi_B - \frac{15 \pi}{8}, \phi_B) \nonumber\\
    C&= (- \phi_C + 2\pi , \phi_C) \nonumber\\
    D&= (-\phi_D + \frac{\pi}{4}, \phi_D) \nonumber\\
    E&= (\phi_A - \phi_C + \pi, \phi_A + \phi_C -\pi) \nonumber\\
    F&= (\phi_B - \phi_D - \frac{29 \pi}{16}, \phi_B + \phi_D - \frac{\pi}{16}),
\end{align}
where $\phi_A, \phi_B, \phi_C, \phi_D$, the angular coordinates for the respective points, are the free parameters in our analysis. 
We find that, to uncover counter-examples to the amended GCW proposal \eqref{prop2}, we must consider situations in which $\mathcal{E}[\hat{\mathcal{W}}_1]$ is strictly larger than the causal wedge $\mathcal{C}[\hat{\mathcal{W}}_1]$, corresponding to an entanglement wedge which includes the defect.
Therefore, together with the usual geometrical constraints, in this setup we have
\begin{align}
    \dfrac{\pi}{8}<\phi_A< \dfrac{15 \pi}{16} \quad &\text{and} \quad \phi_A+\pi<\phi_B< \dfrac{31 \pi}{16} \: ,\\
    \dfrac{31 \pi}{16}<\phi_C< 2\pi \quad &\text{and} \quad \phi_C-\dfrac{15 \pi}{8}<\phi_D< \dfrac{\pi}{8} \: .
\end{align}
Further details regarding the ranges of $\phi$ corresponding to connected versus disconnected phases are included in appendix~\ref{app:phases}.

\subsection{Counter-examples to the GCW}\label{sec:counter}

As advertised in section \ref{sec:proof}, an interesting fact about the entanglement wedges for this chosen setup is that, as opposed to the case of pure AdS$_3$, $\mathcal{E}[\V_{1} \cup \V_{2}]$ being connected does not imply that $\mathcal{E}[\W_{1} \cup \W_{2}]$ is also connected (and vice versa). Indeed, while the (suitably regularized) entropies for their connected candidates match, the entropies for their disconnected ones do not (see appendix \ref{app:phases} for more details). Therefore, it is possible to have $\mathcal{E}[\V_{1} \cup \V_{2}]$ and $\mathcal{E}[\W_{1} \cup \W_{2}]$ both disconnected, both connected, or one connected and the other disconnected.

\begin{figure}[htbp]
    \makebox[\textwidth][l]{
    \begin{minipage}{\textwidth}
        \includegraphics[height=0.4\textheight]{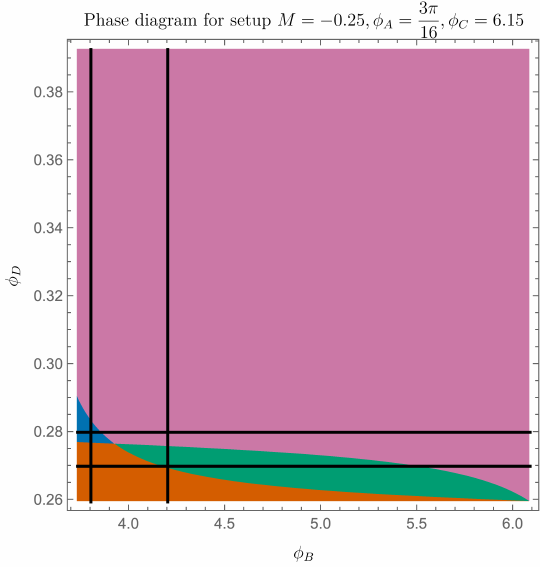}
        \includegraphics[height=0.4\textheight]{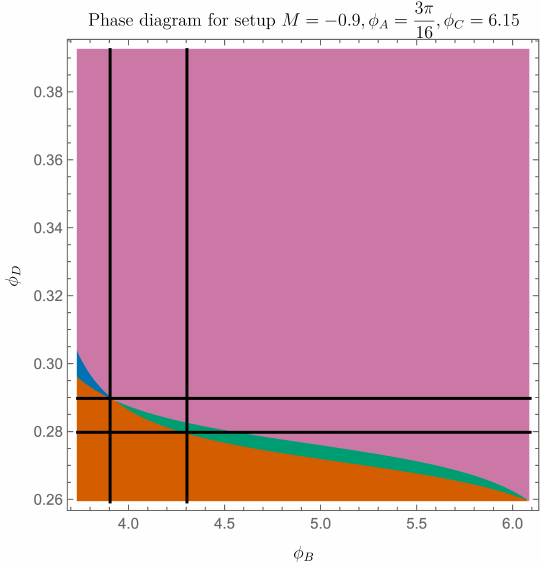}
    \end{minipage}
    }

    \vspace{1em}
    
    \begin{minipage}{\textwidth}
        \centering
        \hspace{-6em}
        \begin{tabular}{|c|c|c|c|c|c|}
            \hline
            $M$ & $\phi_B$ & $\phi_D$ & $\mathcal{E}[\V_{1} \cup \V_{2}]$ & $\mathcal{E}[\W_{1} \cup \W_{2}]$ & $\mathcal{S}_{E}$ \\ [0.5ex] 
            \hline
            -0.25 & 3.8 & 0.27 & disc & disc & nonempty \\
            \hline
            -0.9 & 3.9 & 0.28 & disc & disc & empty \\
            \hline
            -0.25 & 3.8 & 0.28 & disc & conn & nonempty \\
            \hline
            -0.9 & 3.9 & 0.29 & disc & conn & empty \\
            \hline
            -0.25 & 4.2 & 0.27 & conn & disc & nonempty \\
            \hline
            -0.9 & 4.3 & 0.28 & conn & disc & empty \\
            \hline
        \end{tabular}
        \includegraphics[height=0.1\textheight]{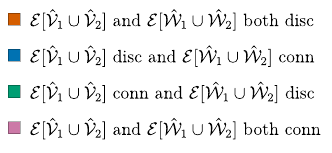}
    \end{minipage}
    \caption{\textbf{Top:} Phase diagrams for the entanglement wedges $\mathcal{E}[\V_1\cup\V_2]$ and $\mathcal{E}[\W_1\cup\W_2]$. \textbf{Bottom:} Counter-examples to any logical relation other than connectedness of both entanglement wedges implying the nonemptiness of $\mathcal{S}_E$.
    The lines in the phase diagrams correspond to the examples displayed in the table.}
    \label{fig:phasediag+table}
\end{figure}

In section \ref{sec:proof} we have shown that having both wedges connected implies $\mathcal{S}_{E} \neq \emptyset$. We now want to show that no other logical relationship can hold. To do so, we consider conical defect spacetimes with two different mass parameters, $M=-0.25$ and $M=-0.9$, and fix the leftmost point of $\W_1$ and $\W_2$, which together guarantee that the leftmost point of $\V_2$ is also fixed (recall that we have fully fixed $\V_1$). (Specifically, we fix $\phi_A=\frac{3\pi}{16}, \phi_C=6.15$.) By varying the rightmost point of $\W_1$ and $\W_2$ --- which implies varying the rightmost point of $\V_2$ accordingly --- it turns out that we can sweep all possible phases for the connectedness of the wedges. We select particular cases for $\phi_B$ and $\phi_D$ to illustrate that for each of the remaining three phases to be considered (i.e. one or both of the entanglement wedges are disconnected), $\mathcal{S}_E$ can be empty or non-empty; see the table in the bottom of figure \ref{fig:phasediag+table}. 
In particular, the first example presented in the table is remarkable: despite having both entanglement wedges in the disconnected phase, their intersection is nonempty. This is a notable counter-example to the conjecture that the non-emptiness of $\mathcal{S}_E$ is equivalent to the connectedness of $\mathcal{E}[\hat{\mathcal{V}}_1\cup\hat{\mathcal{V}}_2]$ and $\mathcal{E}[\hat{\mathcal{W}}_1\cup\hat{\mathcal{W}}_2]$. 
In conclusion, other than what we proved in section \ref{sec:proof}, no other logical relation holds.

\section{Discussion} \label{sec:discussion}

The connected wedge theorem \cite{May:2019yxi, May:2019odp, May:2021nrl} provides an intriguing connection between bulk causal structure and boundary correlation, potentially offering an important window into the information theoretic origin of sub-AdS bulk locality. The theorem highlights necessary conditions for a quantum state to be able to support a particular bulk scattering process, and it remains an interesting open question whether one can formulate criteria which are sufficient for ensuring bulk scattering and which make reference exclusively to the entanglement structure of the CFT state. We have seen in this work that adjusting the statement of the CWT by replacing the usual scattering region $\mathcal{S}$ with the bulk subregion $\mathcal{S}_{E}$ defined in \eqref{eq:def_SE}, as suggested by \cite{Leutheusser:2024yvf} in an effort to relate the ``quantum tasks'' perspective on holography to algebraic properties of large $N$ CFTs, does not lead to a satisfactory theorem-with-converse. Nevertheless, one can show that the emptiness or non-emptiness of $\mathcal{S}_{E}$ is indeed related to information theoretic properties of the CFT state, via the implication
\begin{equation}
    \mathcal{E}[\V_{1} \cup \V_{2}] \: \text{and} \: \mathcal{E}[\W_{1} \cup \W_{2}] \: \text{connected} \: \implies \mathcal{S}_{E} \neq \emptyset \: .
\end{equation}
Moreover, by introducing the alternative bulk subregion $\tilde{\mathcal{S}}_{E}$ defined in \eqref{se2}, we can slightly strengthen this to an if-and-only-if statement 
\begin{equation}
    \mathcal{E}[\V_{1} \cup \V_{2}] \: \text{and} \: \mathcal{E}[\W_{1} \cup \W_{2}] \: \text{connected} \: \iff \tilde{\mathcal{S}}_{E} \neq \emptyset \: .
\end{equation}
Despite following from bulk arguments similar to those responsible for the CWT \cite{May:2019odp}, the replacement of $\mathcal{S}$ by $\mathcal{S}_{E}$ and the ``reversed direction'' of the implication appear to call for a rather different interpretation. Below, we will briefly comment on this interpretation, along with other possible future directions. 

\subsubsection*{QI interpretation}

While the bulk regions $\mathcal{S}$ and $\mathcal{S}_{E}$ are closely related, in the sense that both are defined with reference to the same boundary regions $\V_{i}, \W_{i}$ and satisfy $\mathcal{S} \subseteq \mathcal{S}_{E}$, their operational interpretations are rather different. Indeed, one way of understanding the condition $\mathcal{S}_{E} \neq \emptyset$ is as a statement about 
quantum error correction. 
We will attempt to clarify this below, but we emphasize that, in so doing, we will depart from considering the usual ``non-local quantum computation'' task associated with the boundary setup depicted in figure \ref{fig:CWT}, wherein four agents restricted to their respective causal diamonds $\V_{1}, \V_{2}, \W_{1}, \W_{2}$ cooperate to implement some information processing task. 
Instead, we will consider a task in which Alice, who has access to \textit{both} regions $\mathcal{V}_{1}$ and $\mathcal{V}_{2}$ (or more naturally, to a specific code subspace for these regions), must send a message to Bob, who has access to \textit{both} regions $\W_{1}$ and $\W_{2}$, despite the possibility of an \textit{attack} by some adversary with access to the complementary CFT regions. The doability of the non-local quantum computation task is most readily related to the bulk statement that $\mathcal{S}$ is non-empty, whereas, as explained below, the doability of this alternate encoding/decoding task is most readily related to the bulk statement that $\mathcal{S}_{E}$ is non-empty. 

We would like to consider the setting of subsystem quantum error correction with complementary recovery \cite{Harlow:2016vwg}. We suppose that there is some ``physical'' CFT with bipartite Hilbert space $\mathcal{H}_{\text{CFT}} = \mathcal{H}_{\V_{1} \cup \V_{2}} \otimes \mathcal{H}_{\X_{1} \cup \X_{2}}$, and this Hilbert space hosts a bipartite ``code subspace'' $\mathcal{H}_{\text{code}} = \mathcal{H}_{A} \otimes \mathcal{H}_{E}$, with an (approximately) isometric encoding map $V : \mathcal{H}_{\text{code}} \rightarrow \mathcal{H}_{\text{CFT}}$. This map is such that, for any operator $\tilde{O}_{a}$ acting on $\mathcal{H}_{A}$, there exists an operator $O_{A}$ on $\mathcal{H}_{\V_{1} \cup \V_{2}}$ such that
\begin{equation}
    O_{A} V | \psi \rangle = V \tilde{O}_{a} | \psi \rangle \: , \quad O_{A}^{\dagger} V | \psi \rangle = V \tilde{O}_{a}^{\dagger} | \psi \rangle \quad \text{for all } | \psi \rangle \in \mathcal{H}_{\text{code}} \: .
\end{equation}
The CFT also comes with a time evolution operator $U$. 

\begin{figure}
    \centering
    \includegraphics[width=0.2\textwidth]{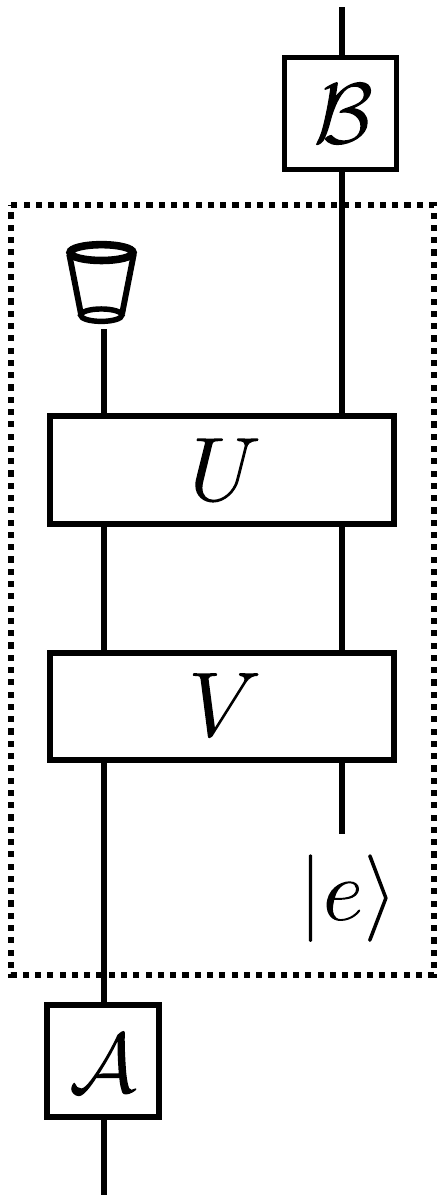}
    \caption{In the setting we consider, Alice and Eve have access to two different subsystems of the code subspace for holographic map $V$. Alice applies a channel $\mathcal{A}$ to encode some message into her share of the state on the code subspace, while Eve chooses her share in an adversarial manner. The holographic map is then applied to the combined state, followed by time evolution via the operator $U$. Part of the resulting state is then traced out. The remaining part is accessed by Bob, who applies an operation $\mathcal{B}$ to attempt to recover Alice's message. Concatenating Eve's input, the isometry $V$, the unitary $U$, and the partial trace into a single quantum channel (dashed box), we can view $\mathcal{A}$ and $\mathcal{B}$ as encoding and decoding operations for this family of error channels. This scenario captures the set-up considered in this paper, where Alice's share corresponds to the bulk state in the entanglement wedge of $\V_{1} \cup \V_{2}$, Eve's share corresponds to the bulk state in the entanglement wedge of $\X_{1} \cup \X_{2}$, the subsystem which is traced over is $\Y_{1} \cup \Y_{2}$, and Bob's subsystem is $\W_{1} \cup \W_{2}$. 
    Whenever we are in a situation with $\mathcal{S}_{E} \neq \emptyset$, there must exist $\mathcal{A}$ and $\mathcal{B}$ allowing for recovery of the state which is correct up to errors exponentially small in $1/G$. }
    \label{fig:error_channel}
\end{figure}

We will now consider a scenario in which Alice has some message $\rho_{M} \in \mathcal{H}_{M}$ that she would like to convey to Bob, despite possible interference from Eve. To do so, she will encode the message into the subsystem $\mathcal{H}_{A}$ of the code subspace via some channel $\mathcal{A} : \mathcal{H}_{M} \rightarrow \mathcal{H}_{A}$ (with script letters now also denoting quantum channels). Before being decoded by Bob, the resulting system is exposed to an error channel which acts as follows. First, Eve adversarially chooses some state $\rho_{e}$ on the subsystem $E$. The holographic map $V$ is then applied to the combined state on $AE$ (which is product by assumption), producing a state in the CFT. The time evolution operator $U$ is then applied to obtain a state on $\W_{1} \cup \W_{2} \cup \Y_{1} \cup \Y_{2}$. Finally, the subsystem $\bar{B} \equiv \Y_{1} \cup \Y_{2}$ is traced out, leaving a state on $B \equiv \W_{1} \cup \W_{2}$. For fixed $V$ and $U$, this results in a family of error channels $\mathcal{E}_{e} : \mathcal{H}_{A} \rightarrow \mathcal{H}_{B}$ dependent on Eve's input state $\rho_{e}$, defined by $\mathcal{E}_{e} (\rho_{A}) = \text{tr}_{\bar{B}}(U(V(\rho_{A} \otimes \rho_{e})))$. After the error channel is applied, Bob may apply his own operation $\mathcal{B} : \mathcal{H}_{B} \rightarrow \mathcal{H}_{M}$ to try to extract the original message. This entire encoding/error/decoding process is depicted in figure \ref{fig:error_channel}.
Requiring $\mathcal{S}_{E}$ to be non-empty is tantamount to the requirement that there is some choice of operations $\mathcal{A}, \mathcal{B}$ for Alice and Bob which allow them to successfully encode and decode the message, at least up to errors which are exponentially small in $1/G$, regardless of which error channel is chosen from the family $\{\mathcal{E}_{e}\}_{e}$. Indeed, when $\mathcal{S}_{E}$ is non-empty, Alice and Bob both have access to this bulk region, whereas Eve is not able to influence the state in this region. 

Our results suggest that Alice and Bob are always able to complete this task, provided the entanglement wedges of $\V_{1} \cup \V_{2}$ and $\W_{1} \cup \W_{2}$ are both connected; this is implicitly a property of the holographic map $V$ and the choice of subsystem which is traced out in the error channel.
A possible intuition is that Alice can use non-local operations to encode the information in a robust way which is impervious to the noise corresponding to Eve's operation followed by time evolution and tracing out the $\Y_{1} \cup \Y_{2}$ subsystem; the error channel can perhaps be thought of as involving some continuum version of a low-depth quantum circuit with local gates. On the other hand, we see that when one of $\mathcal{E}[\V_{1} \cup \V_{2}]$ or $\mathcal{E}[\W_{1} \cup \W_{2}]$ is not connected, it may or may not be the case that the message can be safely transferred; this apparently depends on the details of the state and the regions to which Alice and Bob have access. It is nonetheless worth noting that, in every example where $\mathcal{S}_{E}$ is non-empty, Alice and Bob must both make use of operators outside of their causal wedges: the causal wedge of each is not accessible to the other. Indeed, it is evident that Alice and Bob's causal wedges cannot intersect by the Gao-Wald theorem \cite{Gao:2000ga}, and one can further argue that Alice's causal wedge cannot intersect Bob's entanglement wedge and vice versa \cite{Headrick:2014cta}. In the static examples we consider in section \ref{sec:counter}, this is equivalent to saying that Alice and Bob must use operators which would be identified as ``complex'' according to the python's lunch conjecture \cite{Brown:2019rox, Engelhardt:2021mue}. 

One might wonder whether access to operators outside of the causal wedge by both Alice and Bob, in addition to being necessary, is in fact sufficient for them to communicate safely, but this is not the case: following the analysis in section \ref{sec:counter-ex}, one can identify situations in which both Alice and Bob have access to an entanglement wedge which is strictly larger than their causal wedge, but the region $\mathcal{S}_{E}$ remains empty.\footnote{The cases considered there for which $\mathcal{E}[\V_{1} \cup \V_{2}]$ and $\mathcal{E}[\W_{1} \cup \W_{2}]$ are both disconnected and $\mathcal{S}_{E} = \emptyset$ are an example of this, since the choice of large $\V_{1}$ and $\W_{1}$ ensures that we are always considering situations in which the entanglement wedge is strictly larger than the causal wedge.} 

\subsubsection*{Future directions}

Our analysis invites a number of avenues for further investigation. At the most direct level, there are several natural questions arising from this paper which would be interesting to investigate in follow-up work, including whether analogous results can be established for $n$-to-$n$ scattering and beyond pure states.\footnote{In \cite{Caminiti:2025gyv}, it was shown that even in pure AdS$_{3}$, the usual $n$-to-$n$ version of the connected wedge theorem~\cite{May:2022clu} admits no converse.} In the context of the ordinary connected wedge theorem, both of these extensions have been studied in~\cite{May:2022clu}.

We also note that the entanglement scattering regions studied here will play an important role in the upcoming work~\cite{ap}. While some progress has been made toward a de Sitter version of the connected wedge theorem~\cite{Franken:2024wmh}, a flat space analog of the connected wedge theorem seems illusive due to the degenerate nature of the conformal boundary.  Recently, a proposal for defining entanglement wedges beyond the context of AdS/CFT has been put forward in~\cite{Bousso:2022hlz,Bousso:2023sya}. This construction defines a notion of the entanglement wedge for any bulk region. Besides establishing an analogue of the upper bound~\eqref{eq:lbmi} for general scattering configurations, we will see in~\cite{ap} that we can bound the mutual information in terms of entropies of the ``points-based'' $\cal S$ and entanglement region ${\cal S}_E$ themselves, rather than just areas of the respective ridges.

More broadly, one way of looking at this work is as a reversal of the usual philosophy behind the connected wedge theorem. In the original work \cite{May:2019yxi}, a certain constraint on holographic spacetimes was predicted on the basis of purely quantum information-theoretic arguments, and an explicit geometrical proof was then found in \cite{May:2019odp}. One can reverse this logic: using the proof techniques of the connected wedge theorem, one can proceed to derive various superficially similar but conceptually distinct consequences in the bulk, and we might expect that these should in turn be enforced by some properties of the boundary theory. 
In many cases, these consequences may be mysterious or unexplored from the boundary perspective. In this work, we have found a new implication for the relationship between boundary correlation and causal structure. In the algebraic language, this is the statement that super-additivity of the algebras $\mathcal{A}_{\V_{1} \cup \V_{2}}$ and $\mathcal{A}_{\W_{1} \cup \W_{2}}$ implies a non-empty intersection for these algebras. Understanding this property better from the boundary perspective, or from the information theoretic perspective we have briefly mentioned in this section, is an open question. Similarly, in upcoming work \cite{up}, we will uncover new consequences of scattering for non-minimal extremal surfaces, and we will comment on the interpretation of these consequences in light of conjectures regarding the complexity of the holographic map \cite{Brown:2019rox, Engelhardt:2021mue}. 
Applying this perspective on the connected wedge theorem and its generalizations appears to offer an opportunity to generate a wealth of new insights into the information theoretic properties of holographic quantum theories.  

\section*{Acknowledgments}
We would like to thank Athira Arayath, Levy Batista, Jacqueline Caminiti, Angela Iria Alonso Esteban, Hong Liu, Alex May, and Rob Myers for helpful discussions and/or  comments on the draft. CL thanks Jacqueline Caminiti for sharing the Mathematica notebooks from her paper \cite{Caminiti:2024ctd}, which we have drawn on in section \ref{sec:counter-ex}.
CL acknowledges the support from the Natural Sciences and Engineering Research Council of Canada through a Vanier Canada Graduate Scholarship [Funding Reference Number: CGV -- 192752]. SP is supported by the Celestial Holography Initiative at the Perimeter Institute for Theoretical Physics and by the Simons Collaboration on Celestial Holography. Research
at the Perimeter Institute is supported by the Government of Canada through the Department of Innovation, Science and Industry Canada, and by the Province of Ontario
through the Ministry of Colleges and Universities.

\appendix

\section{Causal ordering of extended ridges} \label{app:vw}

\begin{figure}
    \centering   
    \includegraphics[height=8cm]{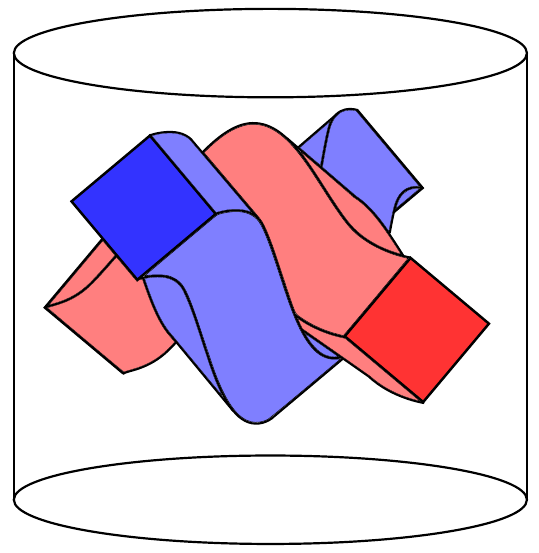}
    \caption{A sketch of the situation ruled out in this appendix, where the entanglement wedges $\mathcal{E}[\V_{1} \cup \V_{2}]$ (red) and $\mathcal{E}[\W_{1} \cup \W_{2}]$ (blue) are connected and non-intersecting, but the latter ``passes under'' the former. }\label{fig:VWWrongOrder}
\end{figure}

We would like to justify the claim from section \ref{sec:proof} that, under the assumptions that 
\begin{enumerate}
    \item $\mathcal{E}[\V_{1} \cup \V_{2}]$ and $\mathcal{E}[\W_{1} \cup \W_{2}]$ are both connected, and 
    \item $\mathcal{E}[\V_{1} \cup \V_{2}]$ and $\mathcal{E}[\W_{1} \cup \W_{2}]$ are non-intersecting (i.e. $\mathcal{S}_{E} = \emptyset$), 
\end{enumerate}
as we assume in the remainder of this section, one has no point $p \in \rho_{\X}^{+}$ to the future of a point $q \in \rho_{\Y}^{+}$. 
As we mentioned in the main text, a non-empty intersection would require the entanglement wedge $\mathcal{E}[\W_{1} \cup \W_{2}]$ to ``pass under'' the entanglement wedge $\mathcal{E}[\V_{1} \cup \V_{2}]$ (see Figure \ref{fig:VWWrongOrder}); we will explain this in more detail, and argue that the latter is not possible. 

In the following, it will be useful to appeal to AdS-hyperbolicity to introduce some time coordinate $t$ and a foliation by AdS-Cauchy slices $\Sigma_{t}$, all of which are topologically equivalent to some $\Sigma_{0}$. We thus have some family of non-intersecting timelike curves $\mu_{p}(t)$ for every $p \in \Sigma_{0}$, which fill $\mathcal{M}$ and satisfy $p = \mu_{p}(0)$ and $\mu_{p}(t) \in \Sigma_{t}$. 
We can define the surjection $\Pi : \mathcal{M} \rightarrow \Sigma_{0}$ which maps from the point $\mu_{p}(t)$ to the point $\mu_{p}(0)$ for any $p, t$. 

\begin{figure}
    \centering   
    \includegraphics[height=8cm]{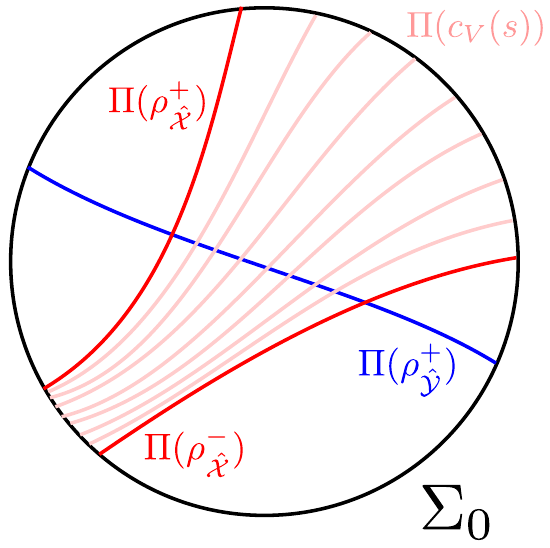}
    \caption{Schematic illustration of the curves $\rho_{\Y}^{+}$ and $c_{V}(s)$ projected onto the slice $\Sigma_{0}$ via $\Pi$. The $c_{V}(s)$ smoothly interpolate between $\rho_{\X}^{+}$ and $\rho_{\X}^{-}$, and the endpoints of $\Pi(c_{V}(s))$ are on opposite sides of $\Pi(\rho_{\Y}^{+})$. }\label{fig:extendedridgefamily}
\end{figure}

First, we will argue that, if $\rho_{\X}^{+}$ had a point in the future of $\rho_{\Y}^{+}$, then $\rho_{\X}^{-}$ would necessarily also have such a point. Suppose for a contradiction that the opposite were true. 
Let $\Gamma_{V}$ denote the set of connected, achronal, codimension-2 surfaces $\gamma_{V} \subset \mathcal{E}[\V_{1} \cup \V_{2}]$ with endpoints in $\V_{1}$ and $\V_{2}$ respectively. 
We can then specify some continuous function
\begin{equation}
    c_{V} : [0, 1] \rightarrow \Gamma_{V} 
\end{equation}
such that
\begin{equation}
    c_{V}(0) = \rho_{\X}^{+} \: , \quad c_{V}(1) = \rho_{\X}^{-} \: ,
\end{equation}
and such that the endpoints of $\Pi(c_{V}(s))$ are always on opposite sides of $\Pi(\rho_{\Y}^{+})$ in $\Sigma_{0}$. See figure \ref{fig:extendedridgefamily}. Here we are using that $\Sigma_{0}$ is assumed to be simply connected, so that $\Pi(\rho_{\Y}^{+})$ divides $\Sigma_{0}$ into two components. 
A consequence is that $\Pi(c_{V}(s))$ and $\Pi(\rho_{\Y}^{+})$ always intersect at at least one point $p$; for the points $p_{1} \in c_{V}(s)$ and $p_{2} \in \rho_{\Y}^{+}$ that get mapped to $p$ by $\Pi$, one must either have $p_{1}$ to the future of $p_{2}$ or vice versa (given that $p_{1} = p_{2}$ is not allowed for $S_{E} = \emptyset$). In particular, it is never the case that $c_{V}(s)$ and $\rho_{\Y}^{+}$ are entirely spacelike separated. 

Now define
\begin{equation}
    T(s) \equiv \sup_{p \in c_{V}(s), \: q \in \rho_{\Y}^{+}} \Delta \tau(p, q)
\end{equation}
to be the maximum (ordered, signed) proper time difference from a point in $\rho_{\Y}^{+}$ to a point in $c_{V}(s)$. 
(This will be well-defined since the curves of interest each contain a point such that the two points are timelike separated.)
This is a continuous function since $c_{V}(s)$ is continuous. Moreover, $T(0) > 0$ and $T(1) \leq 0$ by our assumptions. 
We can then let 
\begin{equation}
    s_{0} \equiv \inf \{ s : T(s) = 0 \} \: .
\end{equation}
The infimum is achieved by continuity and compactness, so $T(s_{0}) = 0$. But one must then have $c_{V}(s_{0})$ and $\rho_{\Y}^{+}$ coinciding at a point (since they cannot be entirely spacelike separated). 
However, this is not possible if $\mathcal{S}_{E} = \emptyset$. This shows that $\rho_{\X}^{+}$ containing a point to the future of $\rho_{\Y}^{+}$ requires $\rho_{\X}^{-}$ to also have such a point.

Now we observe that the condition that $\rho_{\X}^{-}$ has a point to the future of $\rho_{\Y}^{+}$ implies the condition that the ``backward'' scattering region
\begin{equation}
    \mathcal{S}_{Y \rightarrow X} \equiv \mathcal{J}^{+} [ \mathcal{E}[\Y_{1}] ] \cap \mathcal{J}^{+} [ \mathcal{E}[\Y_{2}] ] \cap \mathcal{J}^{-} [ \mathcal{E}[\X_{1}] ] \cap \mathcal{J}^{-} [ \mathcal{E}[\X_{2}] ]
\end{equation}
is non-empty. It therefore remains to show the contrary, i.e. that $\mathcal{S}_{Y \rightarrow X} = \emptyset$. To do this, we will again invoke reasoning similar to that used in \cite{May:2019odp} for the connected wedge theorem \ref{thm:CWT}.

\begin{figure}
    \centering   
    \includegraphics[height=8cm]{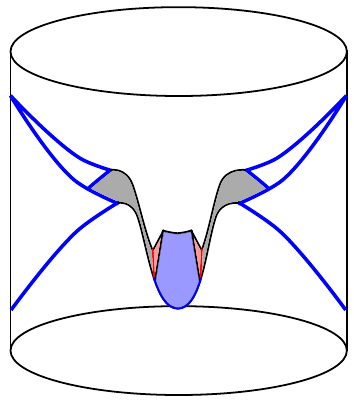}
    \caption{Schematic illustration of the null membrane constructed in the argument for $\mathcal{S}_{Y \rightarrow X} = \emptyset$. The boundary diamonds are the regions $\W_{1}$ and $\W_{2}$, the shaded grey strip is the portion of the slice $\Sigma$ bounded by the RT surface for $\W_{1} \cup \W_{2}$, and the null membrane includes a piece of $\partial \mathcal{J}^{+}[\mathcal{E}[\Y_{1} \cup \Y_{2}]]$ (blue) and a piece of $\partial [\mathcal{J}^{-}[\mathcal{E}[\X_{1}]] \cap \mathcal{J}^{-} [ \mathcal{E}[\X_{2}]]]$ (red). The contradiction surface is obtained by replacing the blue portion of the RT surface with the red portion of the RT surface. }\label{fig:SYX}
\end{figure}

One can again fix a complete achronal slice $\Sigma$, this time containing the RT surface for the region $\W_{1} \cup \W_{2}$, and attempt to construct a null membrane by focusing forward along $\partial \mathcal{J}^{+} [\mathcal{E}[\Y_{1} \cup \Y_{2}] ]$ and backward along $\partial [ \mathcal{J}^{-} [ \mathcal{E}[\X_{1}] ] \cap \mathcal{J}^{-} [\mathcal{E}[\X_{2}] ]]$. Non-emptiness of $\mathcal{S}_{Y \rightarrow X}$ implies that this membrane has the desired topology, with a non-vanishing ridge. A new feature is that the slope will intersect $\Sigma$ at a curve which is not boundary anchored; this is because focusing back from the RT surfaces for $\X_{i}$ can only result in curves with endpoints in the past of $\Sigma$. We thus obtain a null membrane whose intersection with $\Sigma$ is a finite area curve in the bulk, consisting of two segments $y_{1}, y_{2}$ of the RT surfaces for both $\Y_{1}$ and $\Y_{2}$ respectively, and two segments $y_{1}', y_{2}'$ obtained by focusing from $y_{1}, y_{2}$ along the null membrane; see figure \ref{fig:SYX}. Nonetheless, we can form a contradiction surface with smaller area than the putative RT surface by replacing the segments $y_{1}, y_{2}$ in the RT surface with $y_{1}', y_{2}'$, contradicting maximin. It follow that $S_{Y \rightarrow X} \neq \emptyset$ cannot possibly have been true, establishing our claim.  

\section{Conical defect calculations}\label{app:phases}

In this section, we will detail some of the steps for identifying the connected and disconnected phases for $\mathcal{E}[\V_1\cup \V_2]$ and $\mathcal{E}[\W_1\cup \W_2]$ in the conical defect. As discussed in section~\ref{sec:def}, we parametrize the boundary points by $(t, \phi)$. The endpoints are given in equation ~\eqref{eq:phi}, as illustrated in figure~\ref{fig:setupcounterexample}.

Consider a continuous boundary interval. Denote its extreme with a smaller coordinate $\phi$ by the leftmost point, with coordinates $(t_L, \phi_L)$, and the other extreme as the rightmost point, with coordinates $(t_R, \phi_R)$. Let $\Delta t \coloneq t_R - t_L$ and $\Delta \phi \coloneq \phi_R - \phi_L$.\footnote{Since the boundary interval is spacelike, $|\Delta t|< \min (\Delta \phi, 2\pi - \Delta \phi)$, so the difference of cosines is always positive.} Then the length of the smallest Ryu-Takayanagi candidate is

\begin{alignat}{2}
&\log \left[\frac{2 \left[\cos \left(\text{$\Delta $t} \sqrt{|M|} \right)-\cos \left(\Delta \phi  \sqrt{|M|}\right)\right]}{|M| \; \epsilon ^2}\right] \quad &&\text{if} \quad 0 \leq \Delta \phi \leq \pi, \\
&\log \left[\frac{2 \left[\cos \left(\text{$\Delta $t} \sqrt{|M|} \right)-\cos \left((2\pi - \Delta \phi)  \sqrt{|M|}\right)\right]}{|M| \; \epsilon ^2}\right] \quad &&\text{if} \quad \pi \leq \Delta \phi < 2 \pi,
\end{alignat}
where we have introduced a UV cutoff near the boundary at $r=\frac{1}{\epsilon}$.

Consider now the boundary intervals consisting of $\hat{\mathcal{V}}_1 \cup \hat{\mathcal{V}}_2$ and $\hat{\mathcal{W}}_1 \cup \hat{\mathcal{W}}_2$ as defined in section \ref{sec:def}. Each of them may be in either connected or disconnected phases. In our setup (recalling that both $\hat{\mathcal{V}}_1$ and $\hat{\mathcal{W}}_1$ have $\Delta \phi > \pi$ by construction), the entropies for $\hat{\mathcal{V}}_1 \cup \hat{\mathcal{V}}_2$ in the disconnected and connected phases are proportional to
\begin{alignat}{2}
    &s \left( \frac{\pi}{16}, \frac{7 \pi}{8}, \phi_{B} - \phi_{A} - \frac{15 \pi}{16} , \phi_{D} - \phi_{C} + \frac{15 \pi}{8} \right) &&\qquad \text{(disc)} \nonumber\\
    &s \left( \phi_{A} - \frac{\pi}{8} , \frac{31 \pi}{16} - \phi_{B} , \phi_{C} - \frac{31 \pi }{16} , \frac{\pi}{8} - \phi_{D} \right)
    &&\qquad \text{(conn)}
\end{alignat}
where we are denoting
\begin{equation}
    s(a, b, c, d) \equiv \log \left[ \frac{16 \sin(\sqrt{|M|} a) \sin(\sqrt{|M|}b) \sin(\sqrt{|M|} c) \sin(\sqrt{|M|} d)}{|M|^{2} \epsilon^{4}} \right] \: .
\end{equation}

In turn, for $\hat{\mathcal{W}}_1 \cup \hat{\mathcal{W}}_2$, the disconnected and connected phase entropies are proportional to
\begin{alignat}{2}
    &s \left( \frac{\pi}{16} , \frac{\pi}{8}, \phi_{A} - \phi_{B} + \frac{31 \pi}{16} , \phi_{D} - \phi_{C} + \frac{15 \pi}{8} \right) &&\qquad \text{(disc)} \nonumber\\
    &s \left( \phi_{A} - \frac{\pi}{8} , \frac{31 \pi}{16} - \phi_{B} , \phi_{C} - \frac{31 \pi}{16} , \frac{\pi}{8} - \phi_{D} \right) \: .
    &&\qquad \text{(conn)}
\end{alignat}
For a fixed configuration of $\phi$, whichever phase we are in is determined by which of the two (disc) or (conn) are smaller. Notice that the entropies for the connected phases for $\hat{\mathcal{V}}_1 \cup \hat{\mathcal{V}}_2$ and $\hat{\mathcal{W}}_1 \cup \hat{\mathcal{W}}_2$ are equal. However, unlike in pure $AdS$, the disconnected phases do not have the same entropy in general (setups with $\phi_B-\phi_A=\frac{17 \pi}{16}$, for which the angular length for $\hat{\mathcal{W}}_1$ is the same as the one for $\hat{\mathcal{V}}_1$, have disconnected phases with the same entropy). For $\phi_B-\phi_A < \frac{17 \pi}{16}$, the entropy for the disconnected phase for $\hat{\mathcal{V}}_1 \cup \hat{\mathcal{V}}_2$ is smaller than the entropy for the disconnected phase for $\hat{\mathcal{W}}_1 \cup \hat{\mathcal{W}}_2$, and when $\phi_B-\phi_A > \frac{17 \pi}{16}$ the situation is reversed.   In section~\ref{sec:counter} we consider particular configurations that can serve as counter-examples to equation~\eqref{prop2}.

\bibliographystyle{JHEP}
\bibliography{refs}

\end{document}